\documentclass{article}

\pdfoutput=1

\usepackage[utf8]{inputenc}
\usepackage{newtxmath} 
\usepackage{newtxtext} 
\usepackage{fullpage}

\usepackage{authblk}

\usepackage{amsmath}
\usepackage{subfig}
\usepackage{graphicx}
\usepackage[table]{xcolor}
\usepackage{bm}

\usepackage{listings}
\usepackage{tabularx}
\usepackage{upgreek}
\usepackage{subfig}
\usepackage{ulem} 

\usepackage{url}

\rowcolors{2}{gray!15}{}

\newcommand\blfootnote[1]{%
  \begingroup
  \renewcommand\thefootnote{}\footnote{#1}%
  \addtocounter{footnote}{-1}%
  \endgroup
}

\begin{document}

\title{GPU Computing with Python: \\ Performance, Energy Efficiency and Usability}

\author[1,2]{H\aa{}vard~H.~Holm\footnote{Corresponding author: havard.heitlo.holm@sintef.no}}
\author[3,4]{Andr\'{e}~R.~Brodtkorb}
\author[3,4]{Martin~L.~S{\ae}tra}


\affil[1]{
SINTEF~Digital,
Mathematics~and~Cybernetics,
P.O.~Box 124~Blindern,
NO-0314~Oslo,
Norway.}
\affil[2]{
Norwegian~University~of~Science~and~Technology,
Department~of~Mathematical~Sciences,
NO-7491~Trondheim,
Norway.}
\affil[3]{
Norwegian Meteorological Institute, 
P.O.~Box 43~Blindern,
NO-0313~Oslo,
Norway.}
\affil[4]{
Oslo Metropolitan University,
Department of Computer Science,
P.O.~Box 4~St. Olavs plass,
NO-0130~Oslo,
Norway.
}

\date{}

\maketitle

\begin{abstract}
In this work, we examine the performance, energy efficiency and usability when using Python for developing HPC codes running on the GPU. 
We investigate the portability of performance and energy efficiency between CUDA and OpenCL; between GPU generations; and between low-end, mid-range and high-end GPUs. 
Our findings show that the impact of using Python is negligible for our applications, and furthermore, CUDA and OpenCL applications tuned to an equivalent level can in many cases obtain the same computational performance. 
Our experiments show that performance in general varies more between different GPUs than between using CUDA and OpenCL.
We also show that tuning for performance is a good way of tuning for energy efficiency, but that specific tuning is needed to obtain optimal energy efficiency.\blfootnote{This paper is an extended version of our paper published in International Conference on Parallel Computing (ParCo2019)~\cite{holm_parco_2019}}
\end{abstract}


\section{Introduction}
\label{sec:intro}

GPU computing was introduced in the early 2000s, and has since become a popular concept. The first examples were acceleration of simple algorithms such as matrix-matrix multiplication by rephrasing the algorithm as operations on graphical primitives (see e.g.,~\cite{larsen_mcallister_2001}). 
This was cumbersome and there existed no development tools for general-purpose computing. However, many algorithms were implemented on the GPU as proof-of-concepts, showing large speedups over the CPU~\cite{Owens:2007:ASO}. 
Today, the development environment for GPU computing has evolved tremendously and is both mature and stable: Advanced debuggers and profilers are available, making debugging, profile-driven development, and performance optimization easier than ever.

The GPU has traditionally been accessed using compiled languages such as C/C++ or Fortran for the CPU code, and a specialized programming language for the GPU. 
The rationale is often that performance is paramount, and that compiled languages are therefore required. However, for many GPU codes, most of the time is spent in the numerical code running on the GPU. 
In these cases, we can possibly use a higher-level language such as Python for the program flow without significantly affecting the performance. 
The benefit is that using higher-level languages might increase productivity~\cite{7194625}. 
PyCUDA and PyOpenCL~\cite{kloeckner_pycuda_2012} are two Python packages that offer access to CUDA and OpenCL from Python, and have become mature and popular packages since their initial release nearly ten years ago.

Herein, we compare the performance, energy efficiency, and  usability of PyCUDA and PyOpenCL for important algorithmic primitives found in many HPC applications~\cite{Asanovic2006}: a memory bound numerical simulation code and a computationally bound benchmark code. The memory bound code performs simulation of the shallow-water equations using explicit stencils~\cite{brodtkorb2018_nik, gpuocean_testcases_preprint}, and represents a class of problems that are particularly well suited for GPU computing~\cite{HagenHenriksenHjelmervikLie2007, sw12, AMC10:JS, BS12:cmwr,Holewinski:2012:HCG:2304576.2304619}. 
The computationally bound code computes the Mandelbrot set, and we use it to explore limitations of using Python for GPU computing.

We show that accessing the GPU from Python is as efficient as from C/C++ in many cases, demonstrate how profile-driven development in Python is essential for increasing performance for GPU code (up to 5 times), and show that the energy efficiency increases proportionally with performance tuning. 
Finally, we investigate the portability of the improvements and power efficiency both between CUDA and OpenCL and between different GPUs.
Our findings are summarized in tables that justify that using Python can be preferable to C++, and that using CUDA can be preferable to using OpenCL. Our observations should be directly transferable to other similar architectures and problems.

\section{Related Work}
\label{sec:related_work}
There are several high-level programming languages and libraries that offer access to the GPU for certain sets of problems and algorithms. OpenACC~\cite{openacc} is one example which is pragma-based and offers a set of directives to execute code in parallel on the GPU. However, such high-level abstractions are typically only efficient for certain classes of problems and are often unsuitable for non-trivial parallelization or data movement. CUDA~\cite{cuda_programming_guide} and OpenCL~\cite{opencl} are two programming languages that offer full access to the GPU hardware, including the whole memory subsystem. 
This is an especially important point, since memory movement is a key bottleneck in many numerical algorithms~\cite{Asanovic2006} and therefore has a significant impact on energy consumption.

The performance of GPUs has been reported extensively~\cite{brodtkorb2010state}, and
several authors have shown that GPUs are efficient in terms of energy-to-solution. 
Huang et al.~\cite{huang2009energy} demonstrated early on that GPUs could not only speed up computational performance, but also increase power efficiency dramatically using CUDA. 
Qi et al.~\cite{qi2014energy} show how OpenCL on a mobile GPU can increase performance of the discrete Fourier transform by 1.4 times and decrease the energy use by 37\%. 
Dong et al.~\cite{dong2014step} analyze the energy efficiency of GPU BLAST which simulates compressible hydrodynamics using finite elements with CUDA and report a 2.5 times speedup and a 42\% increase in energy efficiency.
Kl\^oh~\cite{kloh2018performance} report that there is a wide spread in terms of energy efficiency and performance when comparing 3D wave propagation and full waveform inversion on two different architectures. They compare an Intel Xeon coupled with an ARM-based Nvidia Jetson TX2 GPU module, and find that the Xeon platform performs best in terms of computational speed, whilst the Jetson platform is most energy efficient. 
Memeti et al.~\cite{memeti2017benchmarking} compare the programming productivity, performance, and energy use of CUDA, OpenACC, OpenCL and OpenMP for programming a system consisting of a CPU and GPU or a CPU and an Intel Xeon Phi coprocessor. They report that CUDA, OpenCL and OpenMP have similar performance and energy consumption in one benchmark, and that OpenCL performs better than OpenACC for another benchmark. In terms of productivity, the actual person writing the code is important, but OpenACC and OpenMP require less effort than CUDA and OpenCL, and CUDA can require significantly less effort than OpenCL.

Previous studies have also shown that CUDA and OpenCL can compete in terms of performance as long as the comparison is fair~\cite{DU2012391, Fang6047190, gimenes_8425194, Karimi2011}, and there have also been proposed automatic source to source compilers that report similar results~\cite{Martinez_6121291, 7832856}.
We do not know of any other established literature that thoroughly compares the performance, usability and energy efficiency of CUDA and OpenCL when accessed from Python.

\section{GPU Computing in Python}
\label{sec:gpu_computing}

In this work, we focus on using Python to access the GPU through CUDA and OpenCL.
These two GPU programming models are conceptually very similar, and both offer the same kind of parallelism primitives. 
The main idea is that the computational domain is partitioned into equally sized subdomains that are executed independently and in parallel.
Even though the programming models are similar, their terminology differs slightly, and in this paper we will use that of CUDA. 
A full review is outside the scope of this work, but can be found in~\cite{opencl_book, cuda_book}. 
The following sections give an overview of important parts of CUDA and OpenCL, and discuss their respective Python wrappers.
We give a short introduction to using them from C++ and Python, and compare the benefits and drawbacks of both approaches.

\subsection{CUDA}
\label{sec:cuda}
CUDA~\cite{cuda_programming_guide} (Compute Unified Device Architecture) was first released in 2007, and is available on all Nvidia GPUs as Nvidia's proprietary GPU computing platform.
It includes third-party libraries and integrations, the directive-based OpenACC~\cite{openacc} compiler, and the CUDA C/C++ programming language. 
Today, five of the ten fastest supercomputers (including number one) use Nvidia GPUs, as well as nine out of the ten most energy-efficient~\cite{web:top500}. 

CUDA is implemented in the Nvidia device driver, but the compiler (\texttt{nvcc}) and libraries are packaged in the CUDA toolkit and SDK.\footnote{Available at \url{https://developer.nvidia.com/cuda-zone}} The toolkit and SDK contain a plethora of examples and libraries.
In addition, the toolkit contains Nvidia Nsight, which is an extension for Microsoft Visual Studio (Windows) and Eclipse (Linux) for interactive GPU debugging and profiling. 
Nsight offers code highlighting, unified CPU and GPU trace of the application, and automatic identification of GPU bottlenecks. 
The Nvidia Visual Profiler is a stand-alone cross-platform application for profiling of CUDA programs, and CUDA versions for debugging (cuda-gdb) and memory checking (cuda-memcheck) also exist.

\subsection{OpenCL}
\label{sec:opencl}
OpenCL~\cite{opencl} (Open Compute Language) is a free and open heterogeneous computing platform that was initiated by Apple in 2009, and today the OpenCL standard is maintained and developed by the Khronos group.
Whilst CUDA is made specifically for Nvidia GPUs, OpenCL can run on a number of heterogeneous computing architectures including GPUs, CPUs, FPGAs, and DSPs. 
The OpenCL API is defined in a common C/C++ header, and a runtime library (ICD loader) redirects OpenCL calls to the appropriate device driver by using an installable client driver (ICD), specific to each architecture.
The actual OpenCL is therefore implemented in each vendor's device driver.  
Contrary to CUDA, there is no common toolkit, but there are several third-party libraries.\footnote{For a list of OpenCL resources, see \url{https://www.khronos.org/opencl/resources}}

Profiling an OpenCL application can be challenging, and the available tools vary depending on your operating system and hardware vendor.\footnote{An extensive list of OpenCL debugging and profiling tools can be found at \url{https://www.khronos.org/opengl/wiki/Debugging_Tools}} 
It is possible to get timing information on kernel and memory transfer operations by enabling event profiling information and adding counters explicitly in your source code.
This requires extra work and makes the code more complex.  
Visual Studio can measure the amount of run time spent on the GPU, and CodeXL~\cite{codexl} can be used to get more information on AMD GPUs.
CodeXL is a successor to gDebugger which offers features similar to those found in Nsight in addition to power profiling, and is available both as a stand-alone cross-platform application and as a Visual Studio extension. 
While it is possible to use Visual Profiler for OpenCL, this requires the use of the command-line profiling functionality in the Nvidia driver, which needs to be enabled through environment variables and a configuration file. After running the program with the profiling functionality in effect, the profiling data can be imported into Visual Profiler. 
Intel Code Builder (part of Intel SDK for OpenCL Applications~\cite{intel_sdk_opencl}) and Intel Vtune Amplifier~\cite{web:intel_vtune_amplifier} can also be used for OpenCL debugging and profiling, but these tools only support Intel CPUs and Intel Xeon Phi processors.

One disadvantage of OpenCL is that there are large differences between the OpenCL implementations from different vendors, and good performance is likely to  rely on vendor-specific extensions. One example is that OpenCL 2.2 is required for using C++ templates in the GPU code, but vendors such as Nvidia only support OpenCL version 1.2. It should also be mentioned that OpenCL has been deprecated in favour of Metal~\cite{web:metal} by Apple in their most recent versions of Mac OS X.

\subsection{GPU Computing from Python}
\label{sec:gpu_from_python}
Researchers spend a large portion of their time writing computer programs~\cite{wilson_2014_best_practices_scicomp}, and compiled languages such as C/C++ and Fortran have been the de facto standard within scientific computing for decades. 
These languages are well established, well documented, and give access to a plethora of native and third-party libraries. 
C++ is the standard way of accessing CUDA and OpenCL today, but developing code with these languages is time consuming and requires great care.
Using higher-level languages such as Python can significantly increase development productivity~\cite{7194625,Prechelt2000}.
However, it should be mentioned that the OpenCL and CUDA kernels themselves are not necessarily made neither simpler nor shorter by using Python: 
The productivity gain comes instead from Python's less verbose code style for the CPU part of the code. 
This influences every part of the host code, from boilerplate initialization code and data pre-processing, to CUDA/OpenCL API calls, post-processing, and visualization of results.

Since Python was first released in 1994, it has gained momentum within scientific computing. Python is easy to learn, powerful, and emphasizes readability of code. All together, this allows for efficient prototyping of code, but unfortunately often at the expense of performance. Since Python is an interpreted language, it does not have the speed of C/C++ and Fortran. For example, a for loop in C/C++ or Fortran will execute at great speed, whilst this construction is relatively slow in Python. This is acceptable for many application areas, but for computationally intensive codes it becomes a showstopper. 

A popular approach to combining the speed of compiled code with Python is to have the program flow written in Python, and the performance critical inner loop in a compiled language. The performance critical code can then be called from Python, using e.g., SWIG wrappers, Numba~\cite{numba_ACM} or Cython~\cite{cython_SciPyProceedings_4}. The following example illustrates this by computing $C = A'A$ with DGEMM from the BLAS library supplied with SciPy:

\begin{minipage}{0.8\linewidth}
\begin{lstlisting}[numbers=left,basicstyle=\ttfamily]
import numpy as np
import scipy.linalg.blas as blas
N = 2048
A = np.ones((N,N))
C = blas.dgemm(alpha=1.0, a=A.T, b=A)
\end{lstlisting}
\end{minipage}

Even though Python can be relatively slow, most of the time in this example is spent in the very efficient BLAS library, giving good overall performance. This approach thus offers the speed of compiled code together with the productivity of a high-level language. We can in a similar way execute GPU code from Python, and our experiments show that this gives good overall performance.
Hence, Python can also be a viable option for production-level code, not only for experimental code and prototypes.

There are several libraries that today offer access to the GPU from Python. One class of libraries are such as OpenCV~\cite{opencv_library} that offers GPU acceleration of algorithms within a specific field. Such libraries are outside the scope of this work, as we focus on general-purpose GPU computing. 
In addition to these types of libraries, Numba~\cite{numba_ACM} and CuPy~\cite{cupy_learningsys2017} are general-purpose programming environments in Python that offer full access to the GPU. However, the GPU programs in Numba are written as Python functions, and the programmer has to rely on Numba for efficient parallelization of the code. While such a design lowers the bar for developers to write code that executes on the GPU, details that are crucial for obtaining the full potential performance might be lost in the abstraction. 
Additionally, Numba is missing support for dynamic parallelism and texture memory.
CuPy also offers functionality to define the GPU functions in terms of Python code, but additionally supports raw kernels written in native CUDA.

PyCUDA and PyOpenCL~\cite{kloeckner_pycuda_2012} are Python packages that offer access to CUDA and OpenCL, respectively.
Both libraries expose the complete API of the underlying programming models, and aim to minimize the performance impact.
The GPU kernels, which are crucial for the inner loop performance, are written in native low-level CUDA or OpenCL, and memory transfers and kernel launches are made explicit through Python functions.
The result is  an environment suitable for rapid prototyping of high-performing GPU code. 
Listing~\ref{listing:pycuda} shows a minimal example for filling an array with the numbers $1$ to $N$. The first four lines import PyCUDA and numpy. Lines six to ten hold the CUDA source code (written in CUDA C/C++), and compiles it using the \texttt{SourceModule} interface. Line eleven retrieves a handle to the kernel so that we can call the GPU function from Python. Line 14 allocates data in Python using numpy, which is then handed over to the GPU kernel in line 15. This manages automatic uploading and downloading of data from the GPU through the \texttt{drv.Out} class. The example shows how PyCUDA can be used to run a GPU kernel in less than 20 lines of code. The equivalent C++ code would be far longer.
\begin{lstlisting}[float,basicstyle={\ttfamily},numbers=left,breaklines=true,label={listing:pycuda},caption={Programming example that shows how to fill an array with the numbers 1 to N using PyCUDA. The corresponding C++ code is much longer.\vspace{1em}}]
import pycuda.autoinit
import pycuda.driver as drv
import pycuda.compiler as compiler
import numpy as np

module = compiler.SourceModule("""
__global__ void fill(float *dest) {
    int i = blockIdx.x*blockDim.x+threadIdx.x;
    dest[i] = i+1;
} """)
fill = module.get_function("fill")

N = 200
cpu_data = np.empty(N, dtype=np.float32)
fill(drv.Out(cpu_data), block=(N,1,1), grid=(1,1,1))
print(cpu_data)
\end{lstlisting}

PyCUDA and PyOpenCL can be installed using popular package managers such as apt-get, pip, and conda. However, the fact that they also require that CUDA and OpenCL are properly installed makes things a bit more complicated. The installation procedure on Windows is especially tricky, and it can be a challenge to install the packages successfully. On Linux, a challenge with OpenCL is that Nvidia only supports version 1.2, whilst PyOpenCL tries to compile with version 2.0 by default when it is installed with pip. This results in runtime crashes when loading the PyOpenCL package. The solution is to manually compile the package, which can be cumbersome. An alternative is to use the packages in apt-get, but these are often very outdated. For PyCUDA, however, we have experienced less difficulties installing on Linux.

PyOpenCL ships with integration for Jupyter Notebooks~\cite{soton403913}, which enables rapid prototyping in an interactive REPL environment\footnote{REPL stands for read-eval-print loop, and is a class of interactive development environments where you execute the code as you write it.}. Using the Jupyter Notebook for prototyping is extremely efficient, and our development cycle has typically been as follows:
\begin{enumerate}
    \item Prototype and develop code in a Jupyter Notebook.
    \item Clean up code in notebook.
    \item Move code from notebook to separate Python modules.
\end{enumerate}
The first stage often entails developing the idea and concept interactively, making test programs, and plotting intermediate and final results. Then, when we have a first implementation, we continue by cleaning up the code and making it suitable for moving into a separate Python module. The tests are added to an appropriate continuous integration environment.

Unfortunately, we have experienced that OpenCL occasionally crashes at random within the Jupyter Notebook environment.
After a significant amount of debugging, we discovered that the context was not properly cleaned up, and variables that went out of scope were not properly deallocated, thereby causing crashes for longer Jupyter sessions. Unfortunately, these crashes demanded a reload of the GPU driver, which typically means rebooting the system.
Our solution was simply to add explicit invocation of Python's garbage collector after dereferencing the variables pointing to GPU main memory.

After having worked with CUDA in the Jupyter Notebook, we discovered that similar problems were also present here. However, because CUDA is somewhat stateful, it uses a stack of contexts and this stack was not properly cleaned up. Our solution was in this case to write a context manager that performed the correct pushing and popping of the CUDA context stack. After having addressed these shortcomings of PyCUDA and PyOpenCL, we have not experienced any crashes caused by the interplay with the Jupyter Notebook.

\subsection{C++ Versus Python}
The GPU can be accessed both from C++ and Python, and these two approaches have their own benefits and drawbacks. In this section we will examine how a computationally bound application -- computing the Mandelbrot set -- can be implemented in the four combinations: C++ and CUDA; C++ and OpenCL; Python and CUDA; and Python and OpenCL.

\begin{figure}
	\begin{centering}
	\includegraphics[width=0.7\textwidth]{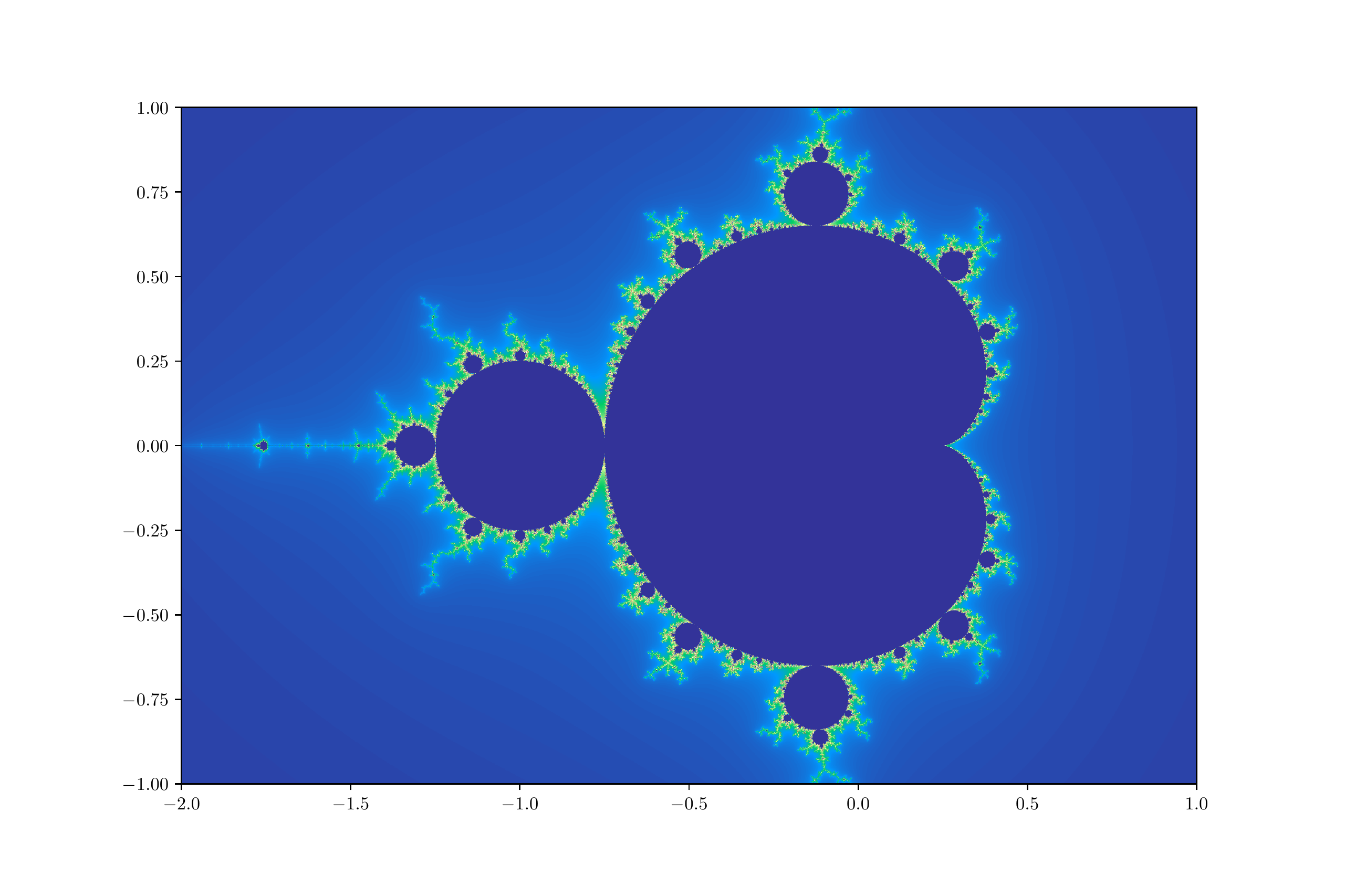}
    \caption{Mandelbrot set colored using continuous coloring. Each complex coordinate $c$ is determined as within the set if the expression $z_{n+1} = z_n^2 + c$ does not diverge after a given set of iterations. The number of iterations performed for each thread is shown here using a continuous color scale. Locations close to the boundary perform many iterations, whilst only a few iterations are required far away. It is evident that threads in close proximity may perform very different number of iterations. }
    \label{fig:mandelbrot}
    \end{centering}
\end{figure}

The Mandelbrot set $M$ consists of all complex points $c$ in which 
\begin{equation*}
	z_{n+1} = z_n^2 + c
\end{equation*}
remains bounded as $n \rightarrow \infty$. In practice, one typically initializes a complex number $c$ and $z_0 = 0$, and iterates until $|z_{n+1}| \geq 2$ or an upper limit of iterations has been reached:
\begin{lstlisting}
//Loop until iterations or until it diverges
while (|z| < 2.0 && n < iterations) {
    z = z*z + c;
    ++n;
}
\end{lstlisting}

Our application consists of a GPU kernel which first initializes the complex coordinate, $c$, based on the position of each thread in the computational grid. Each thread then executes a loop as shown above before the output value for the thread is written to main memory. If we allow e.g., 1000 iterations, a single thread may perform up-to 1000 iterations of the above while-loop before it writes a single number to main GPU memory. However, as Figure~\ref{fig:mandelbrot} shows, the number of actual iterations performed by each thread varies dramatically, even between threads located close to each other. Between CUDA and OpenCL, the kernel code is close to identical, with only syntactic differences (see also Table~\ref{tab:keywords}). The actual kernel that runs on the GPU is the same for both the C++ and the Python variants, and difference between them are therefore found in the host code only.

The host code is responsible for allocating output data on the GPU, launching the kernel, and downloading the result from the GPU to the CPU. To benchmark the different variants, we compute the Mandelbrot set for different extents and measure how much time the CPU and GPU uses for different sections of the code. It should be noted that the overhead of running a traditional Python for-loop is significant, but can be reduced using e.g., so-called list comprehensions. 

Table~\ref{tab:cpp_vs_python} shows a summary of our benchmarks. It shows clearly that the kernel launch overhead is larger for Python than for C++, but it becomes negligible when compared to the kernel execution time. Thus, for kernels that last more than a few ms there will be little performance benefit of using C++, and Python will typically be equally fast. The reason for this is that the CPU in both cases simply will be waiting for the GPU to complete execution for most of the time and the performance difference will be completely masked. This is in particular shown here in the ``Wall time'' row, which shows no practical difference between C++ and Python. In fact, for CUDA the Python variant executes marginally faster. Our explanation to this is that PyCUDA automatically sets the compilation flags for \texttt{nvcc} for the specific GPU, yielding more optimized code. 

\begin{table*}
\caption[]{Performance of CUDA and OpenCL for the Mandelbrot application from both Python and C++. The code was run on an Nvidia Tesla M2090 GPU to compute fifty consecutive zooms onto the point $-0.75 + 0.1i$ with a maximum of 5000 iterations. OpenCL 2.0 is available in the host section of the code, but the device section must still use OpenCL 1.2 for the C++ version. The development time is set subjectively by the authors and the lines of code metric contains only the CPU code related to the actual GPU kernel launch. On Windows we were unable to get asynchronous execution with OpenCL on several different machines, Python versions and GPU driver versions. We have not been able to pinpoint the cause of this, and suspect that asynchronous execution of OpenCL in Windows is not fully supported. We have used page locked memory with CUDA, whilst this was not easily available through OpenCL for the download. The overhead for OpenCL therefore includes the transfer as it cannot be run concurrently with other operations.\footnotemark{} The wall time includes all the time the CPU spends from launching the first kernel to having completed downloading the last result from the GPU.}
\label{tab:cpp_vs_python}
\begin{center}
\begin{tabularx}{0.75\linewidth}{Xrrrr}
\rowcolor{gray!15}
&\multicolumn{2}{c}{\textbf{C++}} & \multicolumn{2}{c}{\textbf{Python}}\\
& \textbf{CUDA} & \textbf{OpenCL} & \textbf{CUDA} & \textbf{OpenCL}\\
API version & 10.0 & 1.2 / 2.0 & 10.0 & 1.2 \\
Development time
& Medium & Medium & Fast & Fast \\
Approximate lines of code & 145 & 130 & 100 & 100 \\
Compilation time & $\sim 5$~s & $\sim 5$~s & $\sim 5$~s & Interactive \\
Kernel launch overhead & 13~$\upmu$s& 318~$\upmu$s
& 19~$\upmu$s & 377~$\upmu$s \\
Download overhead
& 9~$\upmu$s & 4007~$\upmu$s & 52~$\upmu$s & 8872~$\upmu$s \\
Kernel GPU time & 480~ms & 446~ms & 478~ms & 444~ms \\
Download GPU time & 4.0~ms & 3.9~ms & 4.0~ms & 8.8~ms \\
Wall time & 24.2~s & 22.5~s & 24.1~s & 22.7~s \\
\end{tabularx}
\end{center}
\end{table*}
\footnotetext{It should be noted that OpenCL should support efficient pinned memory transfers, but we were unable to get this to work using several driver versions on several different machines. We therefore report the observed transfer times without pinned memory. }

One interesting difference between CUDA and OpenCL from Python is that the compilation time differs. PyCUDA uses \texttt{nvcc} to compile and link an executable in a similar fashion to a regular C++ program. PyOpenCL on the other hand, simply hands over the OpenCL kernel source code to the OpenCL driver which compiles it on-the-fly. We expect that incorporation of the recent Nvidia runtime compilation library (NVRTC) into PyCUDA will alleviate this shortcoming. 

If we compare the other metrics, we see that both the development time and number of lines of code is significantly better for Python. Of particular note is that debugging and visualizing results becomes interactive when using Python and Jupyter Notebooks, thereby increasing the development efficiency dramatically. This is done using standard tools such as Matplotlib~\cite{matplotlib}, making it extremely easy to visualize and explore results. In C++, on the other hand, the only way to reasonably explore the results is through file output (e.g., CSV) and plotting using third-party tools.

\section{Porting, Profiling, and Benchmarking Performance and Energy Efficiency}
We now describe the process of porting code between PyOpenCL and PyCUDA, and optimizating the PyCUDA versions through profile-driven development. 
We consider simulation of the shallow-water equations using three different numerical schemes:
\begin{itemize}
	\item a linear finite difference scheme,
	\item a nonlinear finite difference scheme, and 
	\item a high-resolution finite volume scheme.
\end{itemize}
The schemes are used for simulating real-world ocean currents, and two of them have been used operationally in the early days of computational oceanography.
All three schemes are essentially stencil operations with an increasing level of complexity, and their details are summarized in Holm et al.~\cite{gpuocean_testcases_preprint}.

The numerical schemes are algorithmically well suited for the GPU, but little effort has been made to thoroughly optimize the codes performance on a specific GPU. 
It is well known in the GPU computing community that performance is not portable between GPUs, neither for OpenCL nor CUDA, and automatically generating good kernel configurations is an active research area (see e.g.,~\cite{singh_2017, price_2017, falch_2017}).
We start by porting the three schemes to CUDA, before using the available profiling tools for CUDA to analyze and optimize each scheme.
The obtained optimizations are then also back-ported into the original OpenCL code.
The profiling and tuning is carried out on a laptop with a dedicated GeForce 840M GPU, representing the low-end part of the GPU performance scale, and on a desktop with a mid-range GeForce GTX 780 GPU representing a typical mid-range GPU. 
We compare the performance of the original and optimized implementations with PyCUDA and PyOpenCL using seven GPUs listed in Table~\ref{tab:hardware}, which also includes several high-end server GPUs.

\begin{table*}
\caption{A list of the GPUs used in this work. The profile-driven development was carried out on the 840M and GTX780, and the reminding high-end GPUs were used for performance benchmarking.
We have used the 840M, GTX780, P100 and V100 for benchmarking power efficiency. 
Note that the performance in gigaFLOPS is for single precision. The K80 GPU consists of two processors on the same card and has a boost feature for temporarily increasing the clock speed to increase performance to $2\times 4368$ gigaFLOPS.}
\label{tab:hardware}
\begin{center}
\begin{tabularx}{0.87\linewidth}{XlXrrrr}
	\textbf{Model} & \textbf{Class} & \textbf{Architecture  (year)} &\textbf{Memory}  &  \textbf{GigaFLOPS} & \textbf{Bandwidth} & \textbf{Power device} \\ 
            Tesla \hspace{2em} M2090 & Server & Fermi \hspace{4em} (2011) & 6 GiB  &    1331  & 178 GB/s & N.A.\\ 
            Tesla \hspace{2.4em} K20 & Server & Kepler \hspace{4em} (2012) & 6 GiB  &    3524  & 208 GB/s & N.A.\\   
            GeForce \hspace{0.2em} GTX780 & Desktop & Kepler \hspace{4em} (2013) & 3 GiB &  3977    & 288 GB/s & Watt meter  \\          
            Tesla \hspace{2.4em} K80 & Server & Kepler \hspace{4em} (2014) & $2\times 6$ GiB & $2\times 2795$  &  $2\times240$ GB/s  & N.A.\\
            GeForce \hspace{0.3em} 840M & Laptop & Maxwell \hspace{4em} (2014) & 4 GiB &  790  & 16 GB/s & Watt meter \\  
            Tesla \hspace{2.1em} P100 & Server & Pascal \hspace{4em} (2016) & 12 GiB & 9523 & 549 GB/s & nvidia-smi \\
            Tesla \hspace{2em} V100 & Server & Volta \hspace{4em} (2017) & 16 GiB &  14899 & 900 GB/s  & nvidia-smi\\    
\end{tabularx}
\end{center}
\end{table*}

\subsection{Porting from PyOpenCL to PyCUDA}
The porting process requires changing both the kernel code that runs on the GPU, and the API calls in the CPU code. 
The kernels will in most cases run and produce correct results after a simple change of keywords.
The CPU API calls, however, are quite different between CUDA and OpenCL, and require more attention. This includes handling devices, contexts and streams, compiling and linking kernels, setting kernel arguments, execution of kernels, and memory transfers between the CPU and GPU (or between GPUs). 

A difficulty in the porting process is that it involves a significant amount of changes that all must be completed before it is possible to successfully compile, run and test the code. The Python interpreter and the CUDA compiler can be helpful in the porting process, as they will indicate the locations where code needs to be altered. A summary of key differences between CUDA and OpenCL, which directly correspond to the steps below, can be found in Table~\ref{tab:keywords}. The steps needed to port our simulator from PyOpenCL to PyCUDA should also be applicable for porting other codes:
\begin{enumerate}
  \item Import PyCUDA instead of PyOpenCL.
  \item Change API calls from PyOpenCL to PyCUDA, paying extra attention to context and stream synchronization.
  \item Adjust kernel launch parameters. Block sizes for PyCUDA need to be 3D and global sizes are given in number of blocks instead of total number of threads.
  \item Use CUDA indexing in the kernels. Note that \texttt{gridDim} needs to be multiplied with \texttt{blockDim} to get the CUDA-equivalent of OpenCL \texttt{get\_global\_size()}.
  \item Search and replace the remaining keywords in the kernels. Note that GPU functions in OpenCL have no special qualifier and that GPU main memory pointers need no qualifier for function arguments in CUDA.
\end{enumerate}

\begin{table}
\caption{Keywords, functions and API calls in CUDA and OpenCL. In many cases a simple search and replace is sufficient to translate a program.}
\label{tab:keywords}
\begin{center}
\rowcolors{0}{white}{white}
\ttfamily{}
\setlength{\tabcolsep}{3pt}
\begin{tabularx}{0.7\linewidth}{Xl}
{\normalfont \bfseries CUDA} & {\normalfont \bfseries OpenCL}\\
\rowcolor{gray!15}
\multicolumn{2}{c}{\normalfont Function qualifiers}\\
 \_\_global\_\_ & \_\_kernel\\
 \_\_device\_\_ & N/A\\
\rowcolor{gray!15}
\multicolumn{2}{c}{\normalfont Variable qualifiers}\\
 \_\_constant\_\_ & \_\_constant\\
 \_\_device\_\_ & \_\_global\\
 \_\_shared\_\_ & \_\_local\\
\rowcolor{gray!15}
\multicolumn{2}{c}{\normalfont Indexing}\\
 gridDim & get\_num\_groups()\\
 blockDim & get\_local\_size()\\
 blockIdx & get\_group\_id()\\
 threadIdx & get\_local\_id()\\
 blockIdx*blockDim+threadIdx & get\_global\_id()\\
 gridDim*blockDim & get\_global\_size()\\
\rowcolor{gray!15}
\multicolumn{2}{c}{\normalfont Synchronization}\\
 \_\_syncthreads() & barrier()\\
 \_\_threadfence() & N/A\\
 \_\_threadfence\_block() & mem\_fence()\\
\rowcolor{gray!15}
\multicolumn{2}{c}{\normalfont API calls}\\
kernel\textless{}\textless{}\textless{}...\textgreater{}\textgreater{}\textgreater{}() & clEnqueueNDRangeKernel()\\
 cudaGetDeviceProperties() & clGetDeviceInfo()\\
%
\end{tabularx}
\end{center}
\end{table}

Even though it might be straightforward to port codes between CUDA and OpenCL, there is a large difference in availability of native and third-party libraries. Built-in and specialized functions and data types (e.g., float3) are also different between CUDA and OpenCL (and between Nvidia and AMD). If your code makes use of libraries or built-in functions and data types, the porting process will be more involved as substitutes must be found or implemented.

A difference of practical interest is that CUDA uses the (\texttt{nvcc}) compiler for compilation of the GPU code, whilst OpenCL uses the OpenCL driver. The CUDA compiler is a separate program that performs compilation of both CPU and GPU code and links these together into a single file.
This process can take a significant amount of time, ranging from seconds to minutes depending on how complex the code is. OpenCL, on the other hand, uses the OpenCL driver for compilation, and this process is very fast in our experience. Both CUDA and OpenCL cache their compilations, so this only applies to the first time a kernel is being used. When developing GPU kernels, however, it becomes noticeably slower to work with the compilation times of PyCUDA.

\subsection{Profile-Driven Optimization}
\begin{figure}
	\begin{centering}
    \includegraphics[width=0.9\textwidth, trim=0.85cm 0.0cm 0cm 1.0cm, clip]{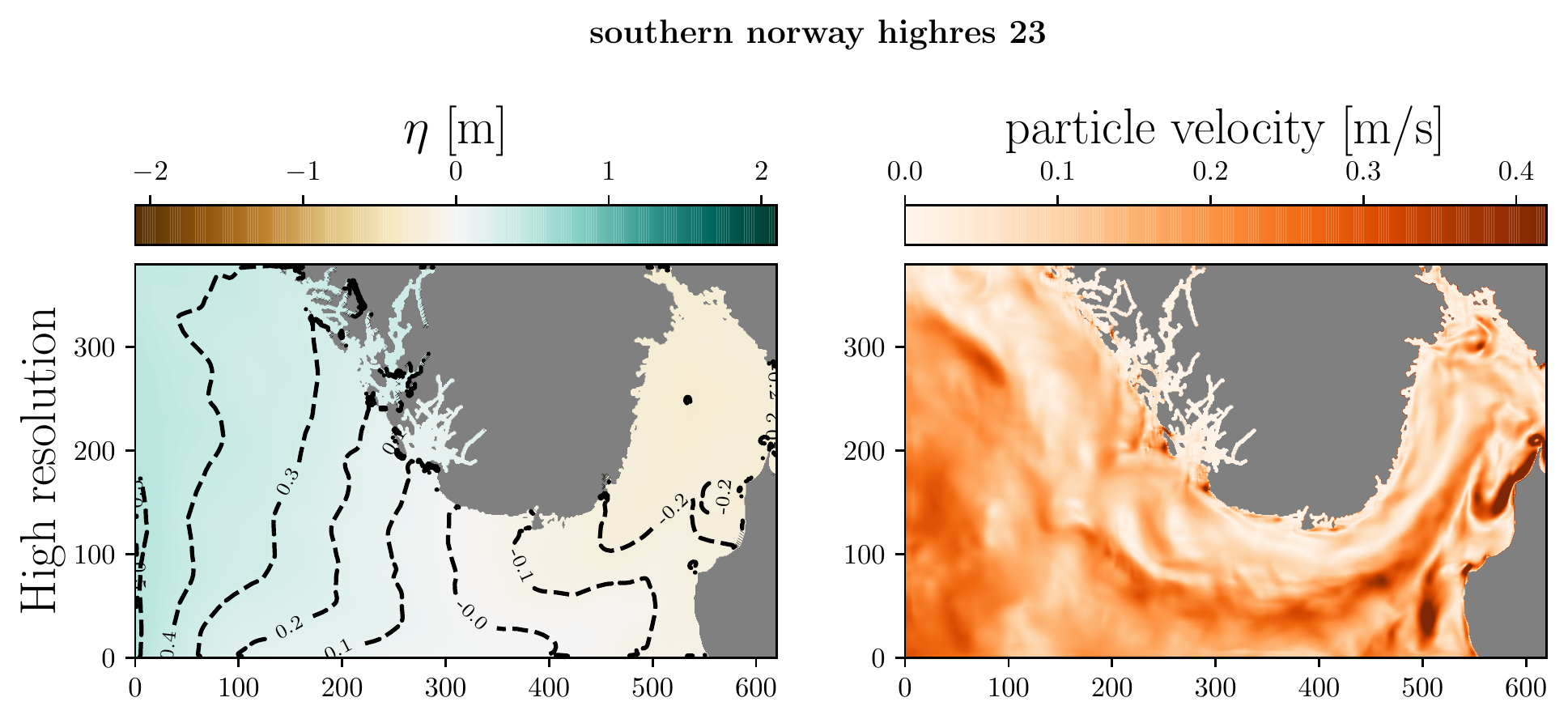}
    \caption{Example of a shallow water simulation for oceanographic purposes with our code. The left figure shows the sea-surface level, whereas the right figure shows the particle velocity. The simulation covers the North Sea around the southern part of Norway, with Denmark in the lower right corner. The axes shows distances in km, and the grid consists of $1550 \times 950$ cells of size $400~\mathrm{m} \times 400~\mathrm{m}$. The simulation here is initialized from operational data provided by the Norwegian Meteorological Institute.}
    \label{fig:shallow_water_simulation}
    \end{centering}
\end{figure}

The shallow-water equations consist of three coupled nonlinear partial differential equations that describe conservation of mass and momentum. 
In our case, they are formulated using $\eta$, the surface deviation from the equilibrium water level, and volume transport $hu$ and $hv$ along the abscissa and ordinate, respectively. Source terms represent varying bathymetry and the Coriolis forces, which takes into account that we solve the equations on a rotating sphere. The equations can be used to model gravitational waves, in which the governing motion is horizontal such as e.g., the ocean~\cite{brodtkorb2018_nik,gpuocean_testcases_preprint}. 
An oceanographic simulation scenario using our simulator is shown in Figure~\ref{fig:shallow_water_simulation}.

We started the analysis of the three different numerical schemes using the Nsight extension in Visual Studio and the standalone Visual Profiler application in the exact same manner as if we were profiling from C/C++. NSight was run on a laptop with a GeForce 840M GPU in Windows, and the Visual Profiler on a desktop with a GTX 780 GPU in Linux. Our workflow started by profiling the code, identifying the performance bottleneck, optimizing the bottleneck, and finally profiling to determine if the optimization was successful~\cite{brodtkorb_etal_13_trends}.
To ensure that our optimizations do not introduce errors in the code, we frequently run integration and regression tests against reference solutions.

\begin{figure}
	\begin{center}
        \hfill
    	\includegraphics[width=0.3\linewidth, trim=0.25cm 3.2cm 1.5cm 1.9cm, clip]{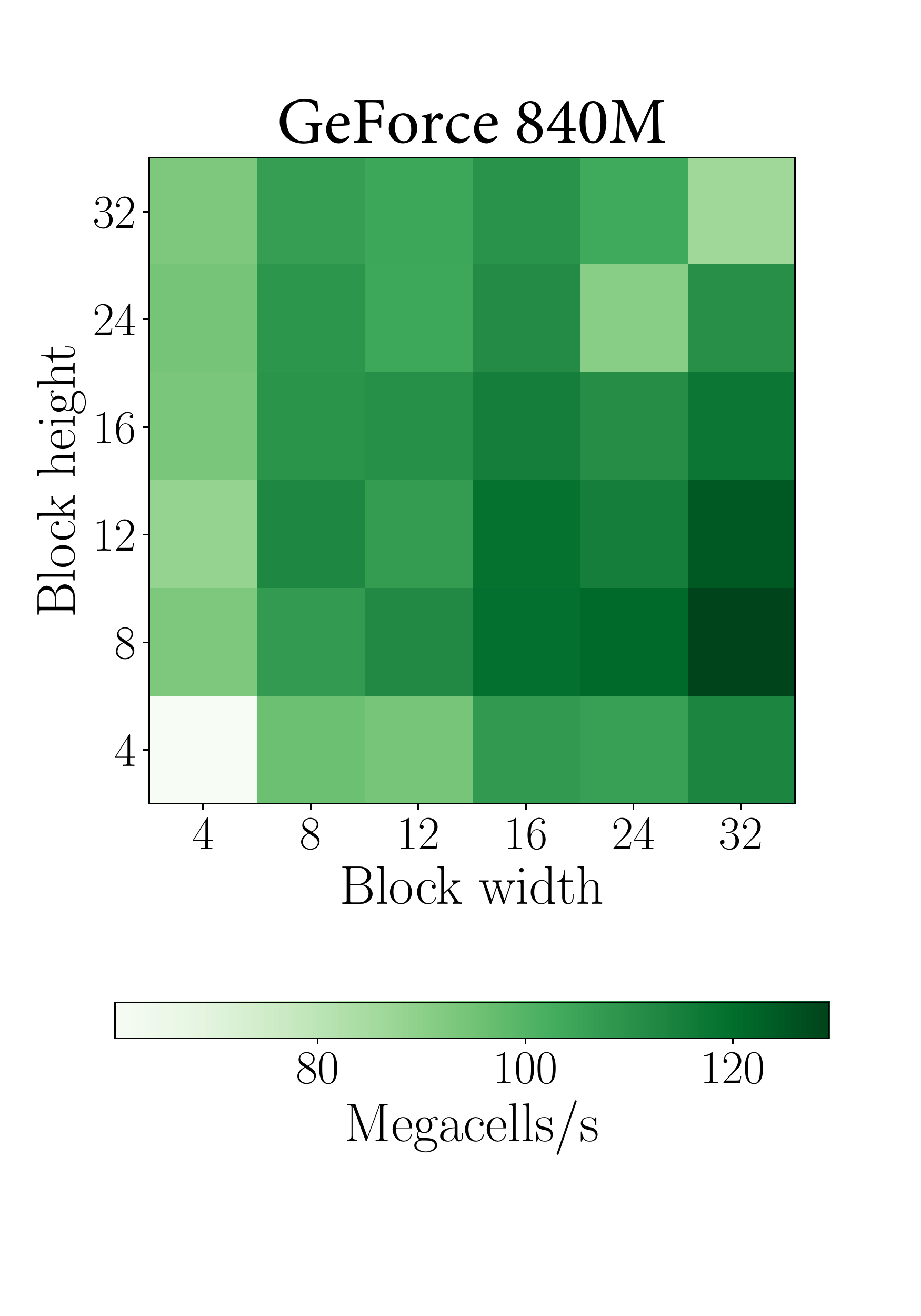}
        \quad
    	\includegraphics[width=0.3\linewidth, trim=0.25cm 3.2cm 1.5cm 1.9cm, clip]{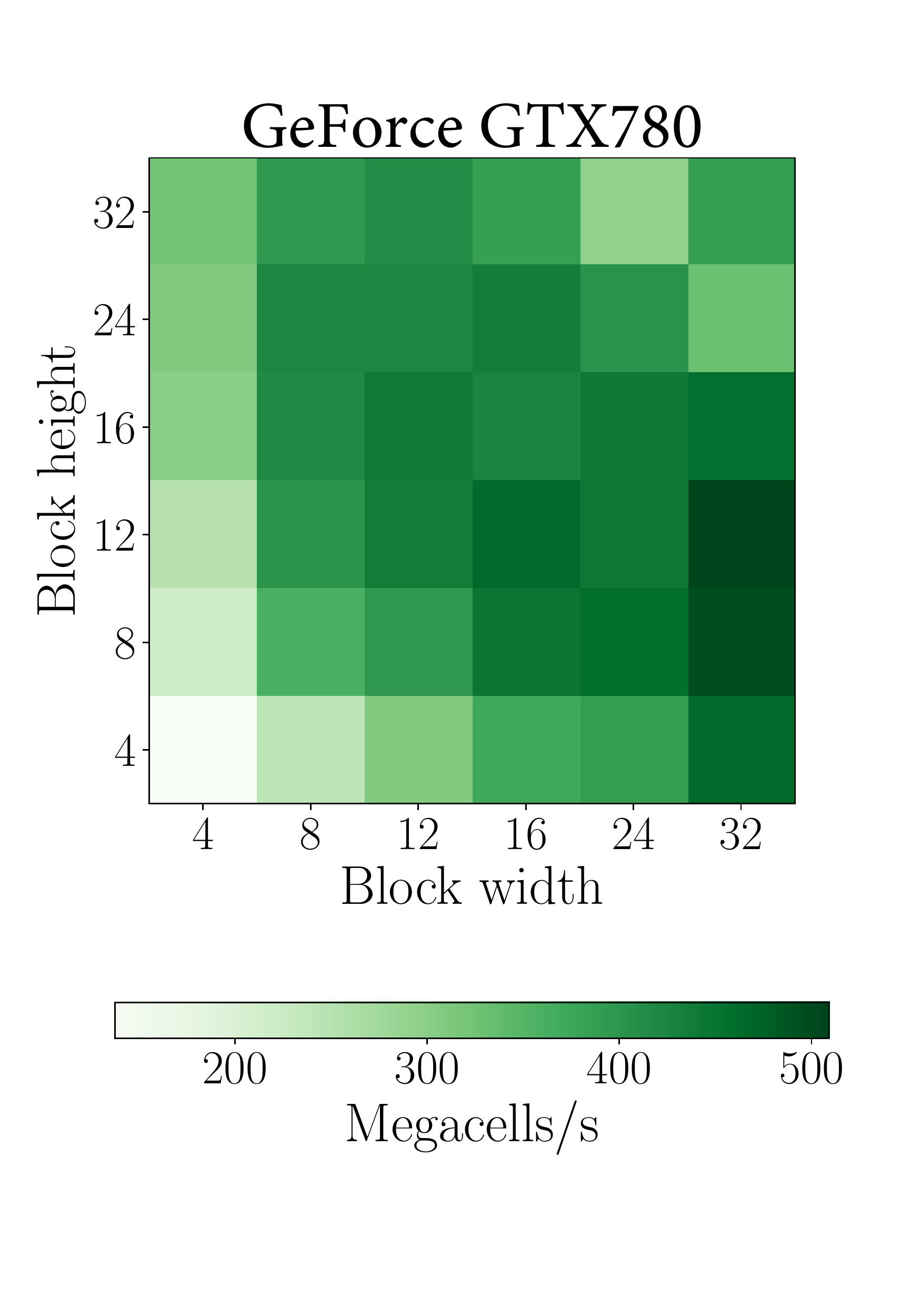}
        \quad
    	\includegraphics[width=0.3\linewidth, trim=0.25cm 3.2cm 1.5cm 1.9cm, clip]{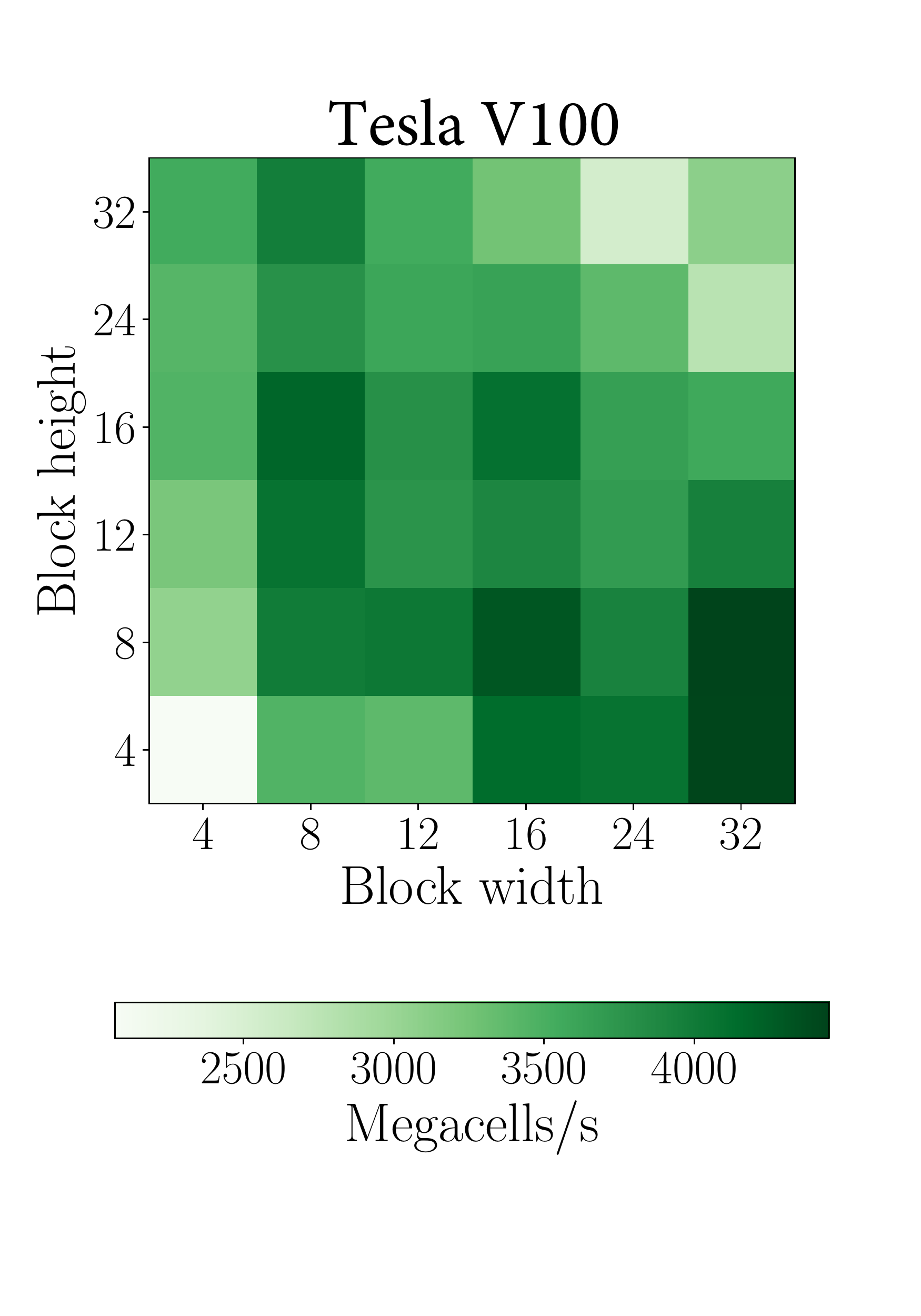}
        \hfill\null
    \caption{Heat map of performance as a function of block width and block height for selected sizes for the high-resolution scheme on three different GPUs. Notice that even though the performance patterns have similarities, the performance on the V100 would be suboptimal if the optimal configuration from the GTX780 is used. The performance increase is 2--5$\times$ for all three GPUs from the slowest to the fastest block size.}
    \label{fig:block_size}
    \end{center}
\end{figure}

An important performance parameter for GPUs is the domain decomposition determined by the block size. CUDA decomposes the work into a grid with equally sized blocks, and all blocks are executed independently. At runtime, the GPU takes the set of blocks and schedules them to the different cores within the GPU. 
Using a too small block size will under-utilize the GPU, and using a too large block size will similarly exhaust the GPU's resources.
Figure~\ref{fig:block_size} shows how the block size has a major impact on performance for three different GPUs, and also illustrates that finding the best block size can be difficult. 
Because of this, we experimentally obtain the optimal configuration for each scheme before starting profiling, and again after the code has been optimized.

\subsubsection{The High-Resolution Finite Volume Scheme}
The first numerical scheme we consider is a high-resolution finite-volume method, which is designed to be well-balanced with respect to steady-state solutions in geostrophic balance~\cite{Chertock2017}. The scheme reads the physical variables, ($\eta, hu, hv$), from GPU main memory, reconstructs an intermediate set of four geostrophic equilibrium variables, calculates inter-cell fluxes, and finally sums up the fluxes and writes the result back to GPU main memory. Each of the steps are stencil operations building on top of each other, and the kernel relies heavily on the use of shared memory for storing and sharing intermediate results. The integration in time is based on a second-order strong stability preserving Runge-Kutta method, which means that the kernel is called twice for every iteration.

The first Nsight analysis of the high-resolution scheme indicates that the occupancy is very low at only 25\%. The occupancy is a measure of how many threads are resident in the cores of the GPU simultaneously, and roughly translates to how well the GPU can hide memory latencies. As the GPU has restricted resources when it comes to the amount of shared memory, number of registers, etc., the occupancy is reduced if each block uses many of these resources. In our case, the limiting factor is the use of shared memory. By reducing the amount of shared memory used by each block, we can expect to get more blocks resident simultaneously on the GPU, which most likely will increase memory throughput and thereby increase performance.

\begin{figure*}
	\begin{center}
	\includegraphics[width=\linewidth, trim=8cm 0.25cm 7cm 0.25cm, clip]{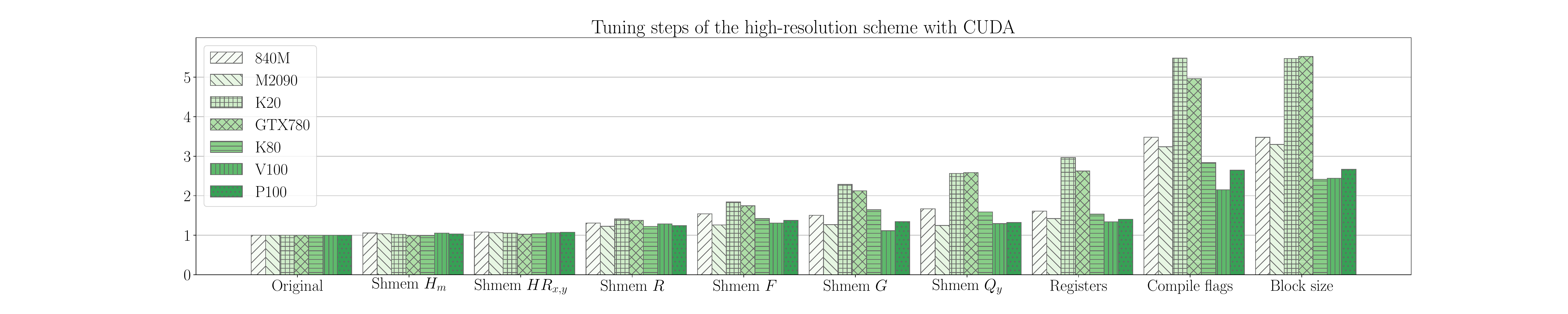}
    \caption{Different optimization stages for the high-resolution kernel showing performance normalized with respect to the original version. Notice that the different optimization strategies impact the performance very differently for different architectures. The platform we profiled on was the 840M GPU, yet the highest performance gain was for the K20 and GTX780 GPUs with over $5\times$ improvement.}
    \label{fig:cdklm_tuning}
    \end{center}
\end{figure*}

Through six consecutive iterations, we progressively reduced the shared memory by 65\% with the following steps (see Figure~\ref{fig:cdklm_tuning}):
\begin{enumerate}
    \item Recomputing bathymetry in cell intersections instead of storing $H_m$.
    \item Recomputing face bathymetry instead of storing $RH_x$ and $RH_y$.
    \item Reusing buffer for physical variables $Q$ for storing the reconstruction variables $R$.
    \item Recomputing fluxes along the abscissa instead of storing $F$.
    \item Recomputing fluxes along the ordinate instead of storing $G$. 
    \item Reusing the buffer for derivatives along the abscissa, $Q_x$, and derivatives along the ordinate, $Q_y$.
\end{enumerate}
In the third and sixth steps we managed to reuse other shared memory buffers by reordering the execution flow of the code, and for all other cases we relied on re-computation. This essentially means that we prefer recomputing results rather than storing and sharing them between threads, thus trading extra computation for less memory storage. 

The first seven groups in Figure~\ref{fig:cdklm_tuning} show the impact of the shared memory on performance. 
The first few buffers we removed were insufficient to increase the occupancy, because it would not free enough space for another resident block per core. However, after stage three there was space for one extra block, and the occupancy increased. After all six iterations occupancy increased to 56\%, memory bandwidth utilization increased by a factor 1.8, and gigaFLOPS more than doubled. It should be noted that memory throughput is the important factor here, both since our kernel is memory bound and because our re-computations artificially increase the number of floating point operations performed by the kernel by 20\%.

The next limiting factor was the number of registers (variables) used by the kernel. First, we removed all variables related to debugging, and packed four boolean kernel arguments into a single integer using bit operations. Finally, we removed several temporary variables by aggregating these as soon as possible into the final computed flux. The compiler is already very good at optimizing register use, but using the above optimizations reduced the number of registers per thread from 49 to 47, which was sufficient to increase occupancy from 56\% to 62\%.

Compilation flags can also be used to increase performance with relatively little effort, as can be seen in Figure~\ref{fig:cdklm_tuning}. Here, we used the \texttt{----use\_fast\_math} flag which enables using fast, albeit less precise, mathematical functions for operations such as exponential and square roots. This gave a dramatic effect and increased memory throughput by a factor of 3.9 relative to the original version. This is not only because the mathematical functions execute faster, but perhaps more importantly because the fast mathematical functions use less register space, reducing from 47 to 40 registers per thread. This increased occupancy to 69\%, and left shared memory as the bottleneck once again. In our case, the use of fast mathematical functions is sufficiently accurate, but these compilation flags can ruin the correctness of a program and should be used with care. Other important flags to consider are flags that determine the maximum amount of register and specify cache configurations. After having optimized the kernel, our block size is probably no longer the optimal, and we rerun the block size optimization. 

The performance increased by a factor 3.5 on the 840M, and 5.5 on the GTX 780 after these optimization steps. One surprising point is that the optimization steps have very different effect on different architectures. 
For example, the K20 and GTX780 increased to over 5 times the performance, whereas the K80, P100 and V100 increased by only half of that.
Some steps also actually decrease the performance on one GPU, whilst having a positive effect on others. This was especially noticeable with our attempts at optimizing register use.
It should still be noted that none of the optimizations steps that reduced shared memory usage led to a slowdown on any of the GPUs, even though these steps increased the computational workload per thread.

\subsubsection{The Nonlinear Finite Difference Scheme}

Our second scheme is a nonlinear second-order classical leapfrog scheme, also called centered-in-time centered-in-space. In this scheme, the three physical variables are defined on a staggered grid, and are updated using separate kernels. Each kernel reads all three variables from GPU main memory, performs computations, and writes back the result. The computational work in these three kernels is significantly less than for the high-resolution scheme, making this scheme even more memory bound. However, this also means that only four shared memory buffers are needed (three for the variables from the current time step and one for the constant depth). The initial profiling shows that the occupancy is 100\% for all three kernels. 

\begin{figure}
	\begin{center}
	\subfloat%
    	{\setlength{\fboxsep}{0pt}\setlength{\fboxrule}{0.1pt}\fbox{%
        \includegraphics[width=0.6\linewidth, trim=0cm 2pt 0cm 3pt, clip]{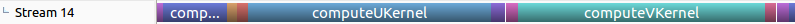}}}  \\
        \vspace*{0.2cm}
	\subfloat%
    	{\setlength{\fboxsep}{0pt}\setlength{\fboxrule}{0.1pt}\fbox{%
        \includegraphics[width=0.6\linewidth, trim=0cm 2pt 0cm 2pt, clip]{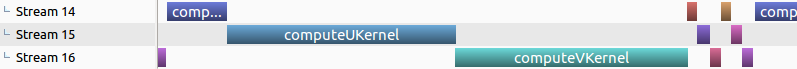}}}\\ 
        \vspace*{0.2cm}
	\subfloat%
    	{\setlength{\fboxsep}{0pt}\setlength{\fboxrule}{0.1pt}\fbox{%
        \includegraphics[width=0.6\linewidth, trim=0cm 2pt 0cm 3pt, clip]{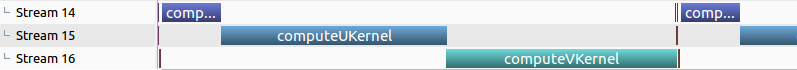}}}\\
        \vspace*{0.2cm}
    \subfloat%
    	{\setlength{\fboxsep}{0pt}\setlength{\fboxrule}{0.1pt}\fbox{%
        \includegraphics[width=0.6\linewidth, trim=0cm 3pt 0cm 2pt, clip]{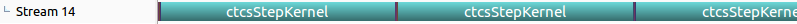}}} 
    \caption{Four screenshots from the Nvidia Visual Profiler, illustrating the effect of optimizations for the nonlinear scheme. The top figure shows the original performance, and in the second we introduce multiple streams so that the three physical variables can be computed independently. In the third figure we reduce the number of blocks that are needed for the boundary condition kernels, and finally we compute all three variables within one kernel in the bottom figure. The horizontal time scale is about 7~ms and equal for all four figures, and shows that computing all variables in a single kernel gives the highest performance gain.}
    \label{fig:ctcs_timeline}
    \end{center}
\end{figure}

Due to its simplicity, this type of kernel leaves less headroom for optimization than the aforementioned high-resolution scheme. For example, the compilation flag \texttt{----use\_fast\_math}, which had a dramatic effect for the high-resolution scheme, has only a marginal effect on this kernel because it performs less mathematical operations and uses few registers already. One thing that we can optimize, however, is to use concurrent kernel execution, since all three variables can be updated simultaneously. Unfortunately, each kernel already occupies all available resources on the GPU, and therefore the effect turns out to be marginal, as shown in Figure~\ref{fig:ctcs_timeline}. The figure also shows that the boundary condition kernels took too long time, and directed our attention to them. These could with relative ease be optimized to take an insignificant amount of time.

Because we launch three kernels to update the variables, we read the variables $\eta$, $hu$ and $hv$ multiple times from GPU main memory for every iteration. By carefully gathering these kernels into one, we can reduce the number of variables read and written from GPU main memory from 16 to ten. After this optimization, our profiling shows a 36\% decrease in memory traffic, compared to the theoretical 37.5\%. This directly translates to an equivalent improvement in performance, as shown in the lower panel of Figure~\ref{fig:ctcs_timeline}.

By merging the three kernels into one, the occupancy decreased to 62.5\%, as each thread now requires more registers. Several attempts were made to reduce this number, but none of our attempts helped the compiler to do a better job. One way of forcing fewer registers is to set the \texttt{----maxrregcount} compilation flag. However, this implies that registers are spilled to local memory (cache on the GPU), and even though the occupancy increases, the actual performance decreases. The final results from optimizing the nonlinear scheme is shown in Figure~\ref{fig:total_results}, and the optimizations increase performance of the CUDA versions by between two to three times on all seven GPUs.

\subsubsection{The Linear Finite Difference Scheme} 

The final scheme solves the linearized shallow-water equations using a forward-backward linear finite-difference scheme~\cite{sielecki68} and is asymmetric in time. It consists of three simple kernels, in which the most recent results are always used (similarly to Gauss-Seidel iteration):
\begin{equation*}
	\begin{split}
    	hu^{n+1} &\leftarrow F(\eta^n, hv^n, hu^n), \\
    	hv^{n+1} &\leftarrow G(\eta^n,  hv^n, hu^{n+1}), \\
    	\eta^{n+1} &\leftarrow H(\eta^n, hv^{n+1}, hu^{n+1}).
	\end{split}
\end{equation*}

In order to reduce the amount of memory read from GPU main memory, we apply the same strategy as for the nonlinear scheme and carefully combine all three kernels into one. Since the execution order has to comply with the data dependencies, we need to read extra input data (referred to as ghost cells or computational halo) for all variables at time step $n$. We then compute $hu^{n+1}$, $hv^{n+1}$, and finally $\eta^{n+1}$.
Our profiling shows that this reduced the amount of memory read and written to GPU main memory by 50\%. 

The profiling now tells us that memory dependencies are the main bottleneck, and that the GPU has no eligible instructions for about 60\% of the cycles. Without a major redesign of the algorithm, there is typically little we can do to improve the kernel further, but for completeness we added the \texttt{----use\_fast\_math} compilation flag. However, since the kernel is simple and heavily memory bound the flag had a negligible impact.

The overall performance increase is more than two times for the laptop 840M GPU and the server V100 GPU, but only around 40\% for the others. It should also be noted that this kernel is the one with the smallest negative performance impact when moving from OpenCL to CUDA.

\subsection{Backporting Optimizations to OpenCL}
To do a fair comparison of the performance of PyCUDA and PyOpenCL, we need to back-port our optimizations to PyOpenCL. As most of the optimizations are carried out in the CUDA source code, it is a relatively simple matter of copying the optimized code into the original OpenCL kernels, and update the code according to Table~\ref{tab:keywords}. In the Python code, the loading of the separate kernels for the linear and nonlinear schemes are replaced by the new merged kernels, and the input arguments are updated to correspond to these optimizations. Active and appropriate use of a version control system and regression testing is crucial in this work.

Porting compiler flags can in general be a greater challenge, as there are no one-to-one overlap between the compile-time options for the two languages. However, in our case only the \texttt{----use\_fast\_math} compilation flag is used in the final PyCUDA version, and it has a matching counterpart, \texttt{-cl-fast-relaxed-math}, that is passed to PyOpenCL's API when compiling the kernels.

\subsection{Comparing Performance}

\begin{figure*}
	\begin{center}
    \null\hfill
	\subfloat{\includegraphics[width=0.45\linewidth, trim=2cm 0.5cm 2cm 0.5cm, clip]%
    	{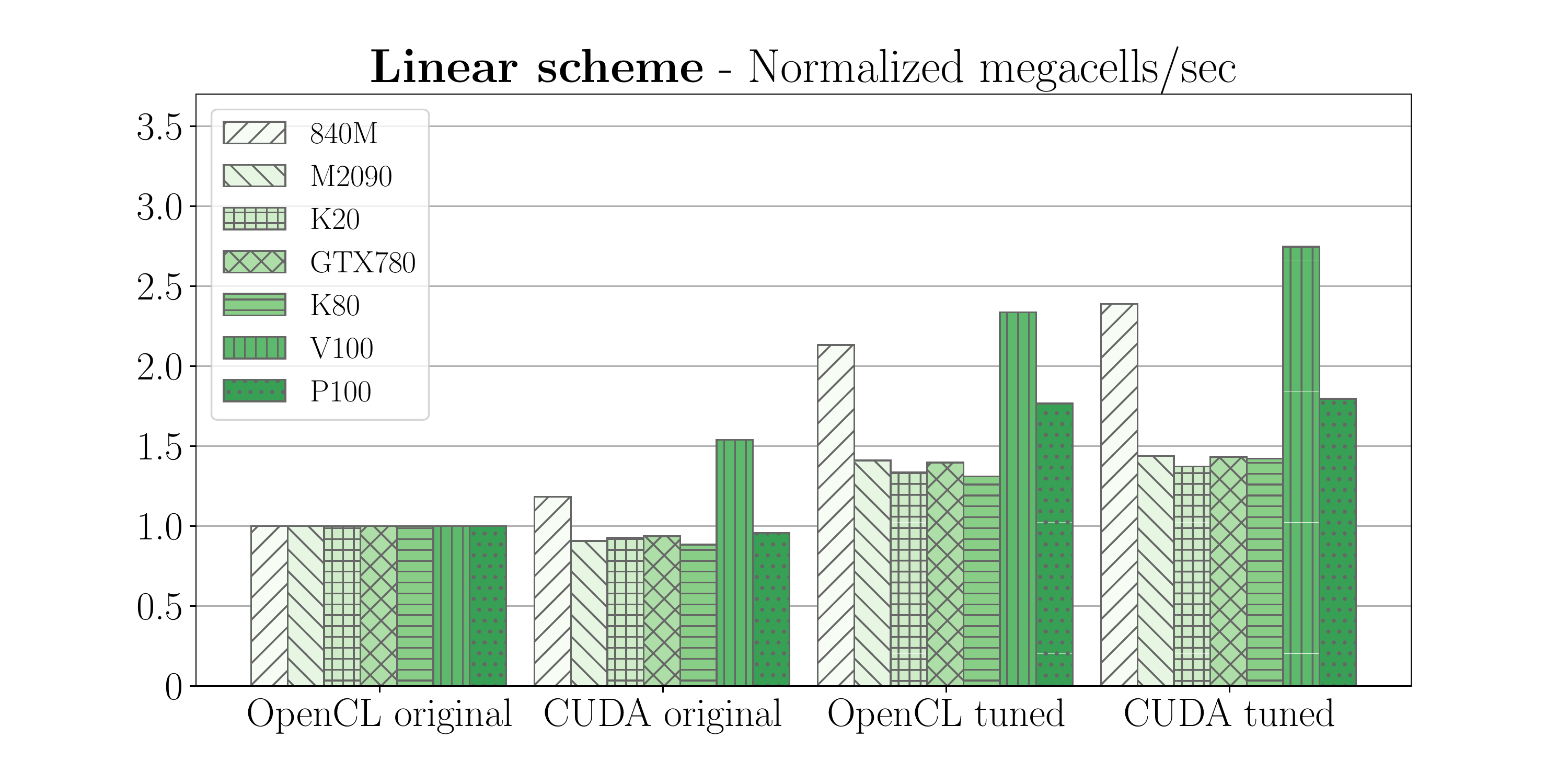}} %
    \qquad%
	\subfloat{\includegraphics[width=0.45\linewidth, trim=2cm 0.5cm 2cm 0.5cm, clip]%
    	{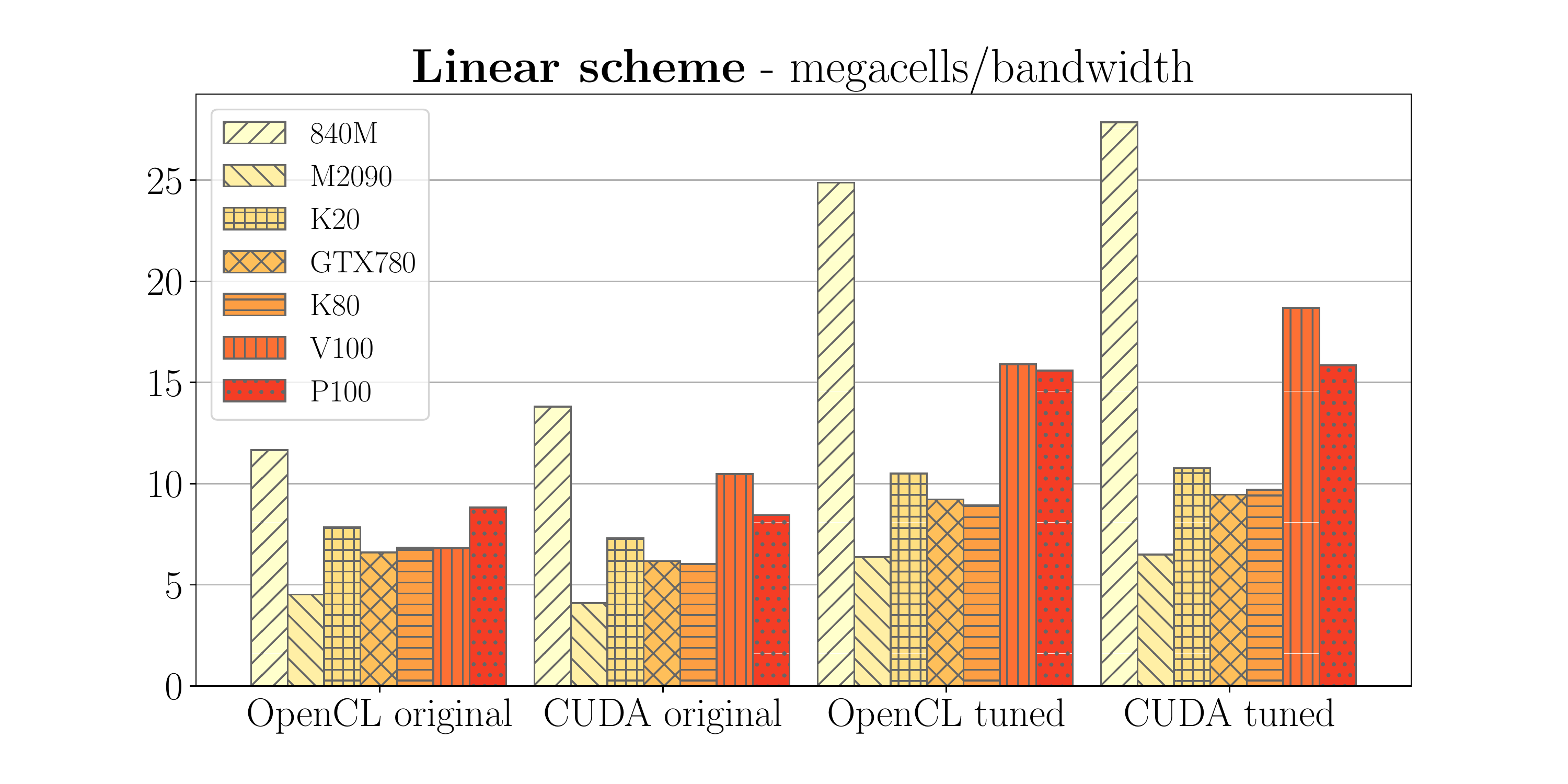}} %
    \hfill\null\\ \vspace{-2em}
    \null\hfill
 	\subfloat{\includegraphics[width=0.45\linewidth, trim=2cm 0.5cm 2cm 0.5cm, clip]%
    	{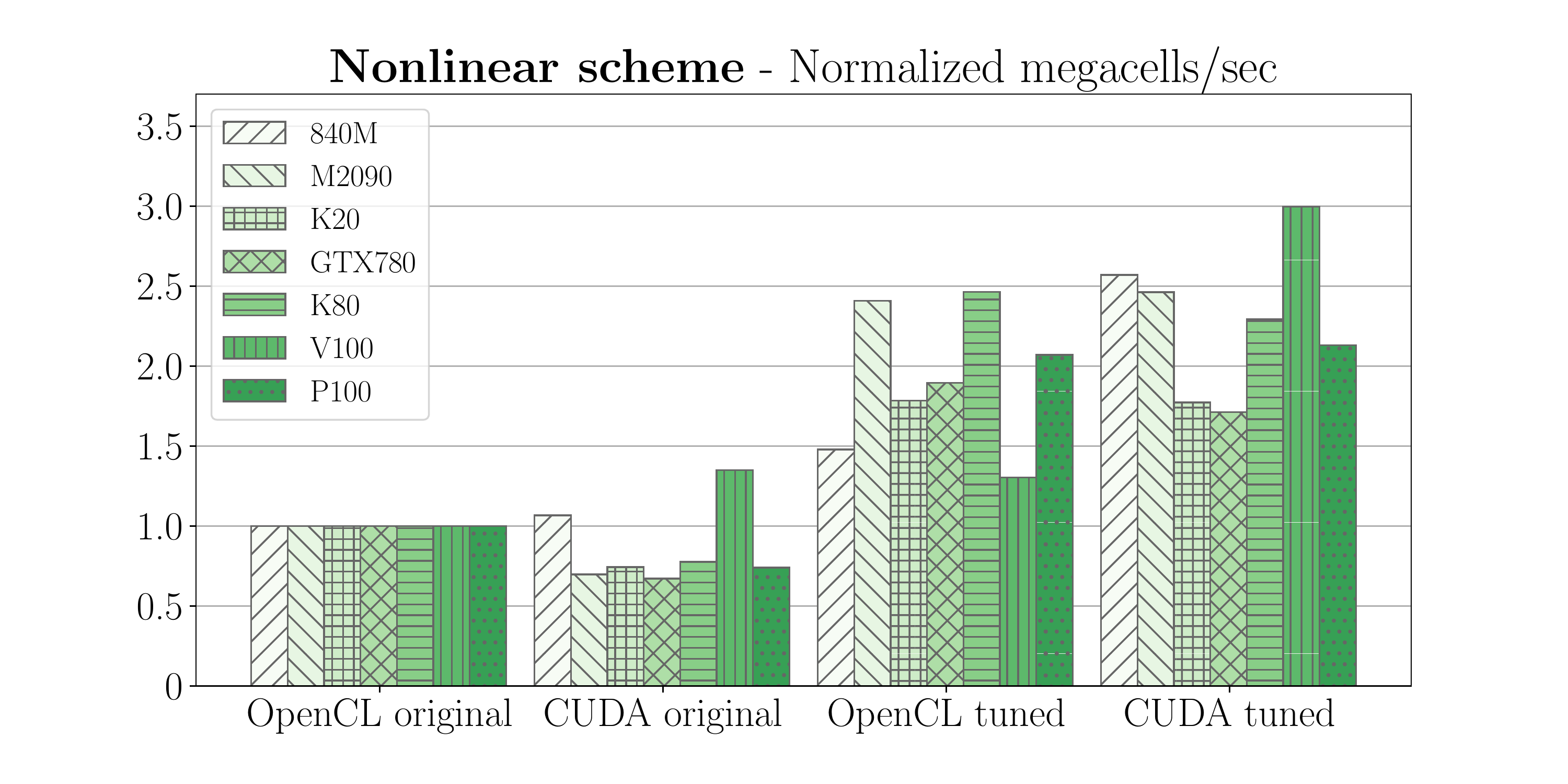}} %
    \qquad%
 	\subfloat{\includegraphics[width=0.45\linewidth, trim=2cm 0.5cm 2cm 0.5cm, clip]%
    	{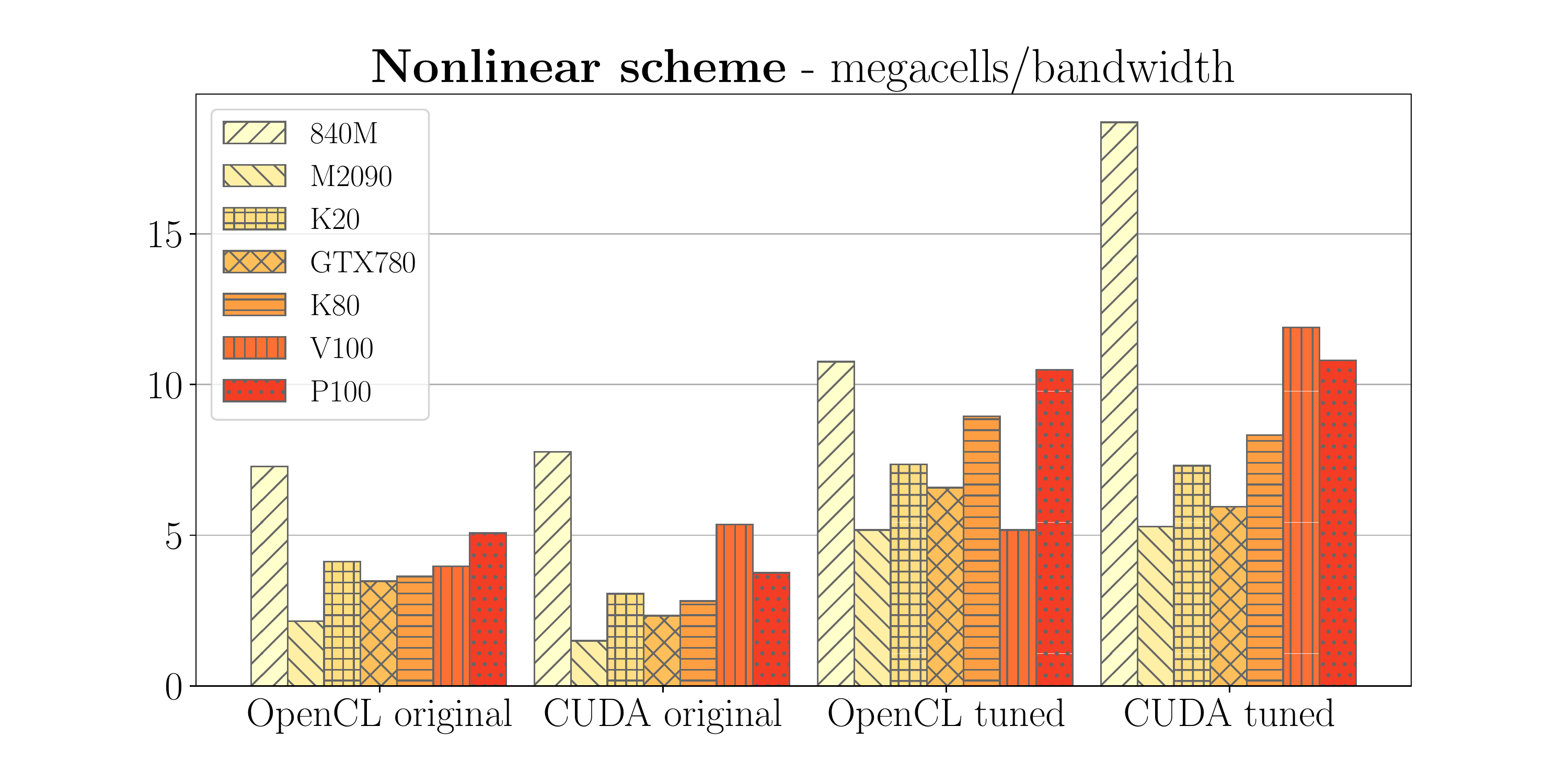}} %
    \hfill\null\\ \vspace{-2em}
    \null\hfill
	\subfloat{\includegraphics[width=0.45\linewidth, trim=2cm 0.5cm 2cm 0.5cm, clip]%
    	{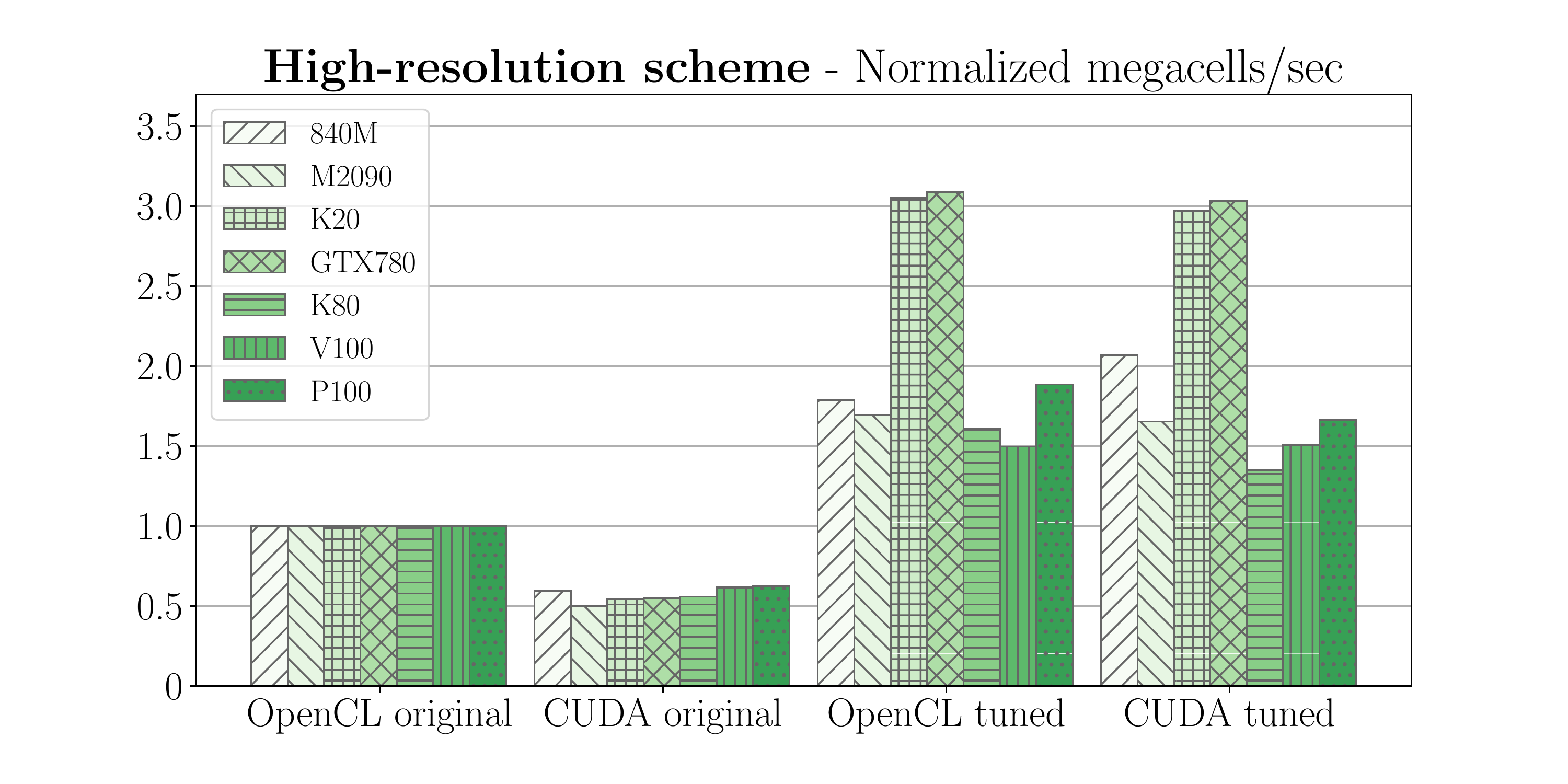}} %
    \qquad%
	\subfloat{\includegraphics[width=0.45\linewidth, trim=2cm 0.5cm 2cm 0.5cm, clip]%
    	{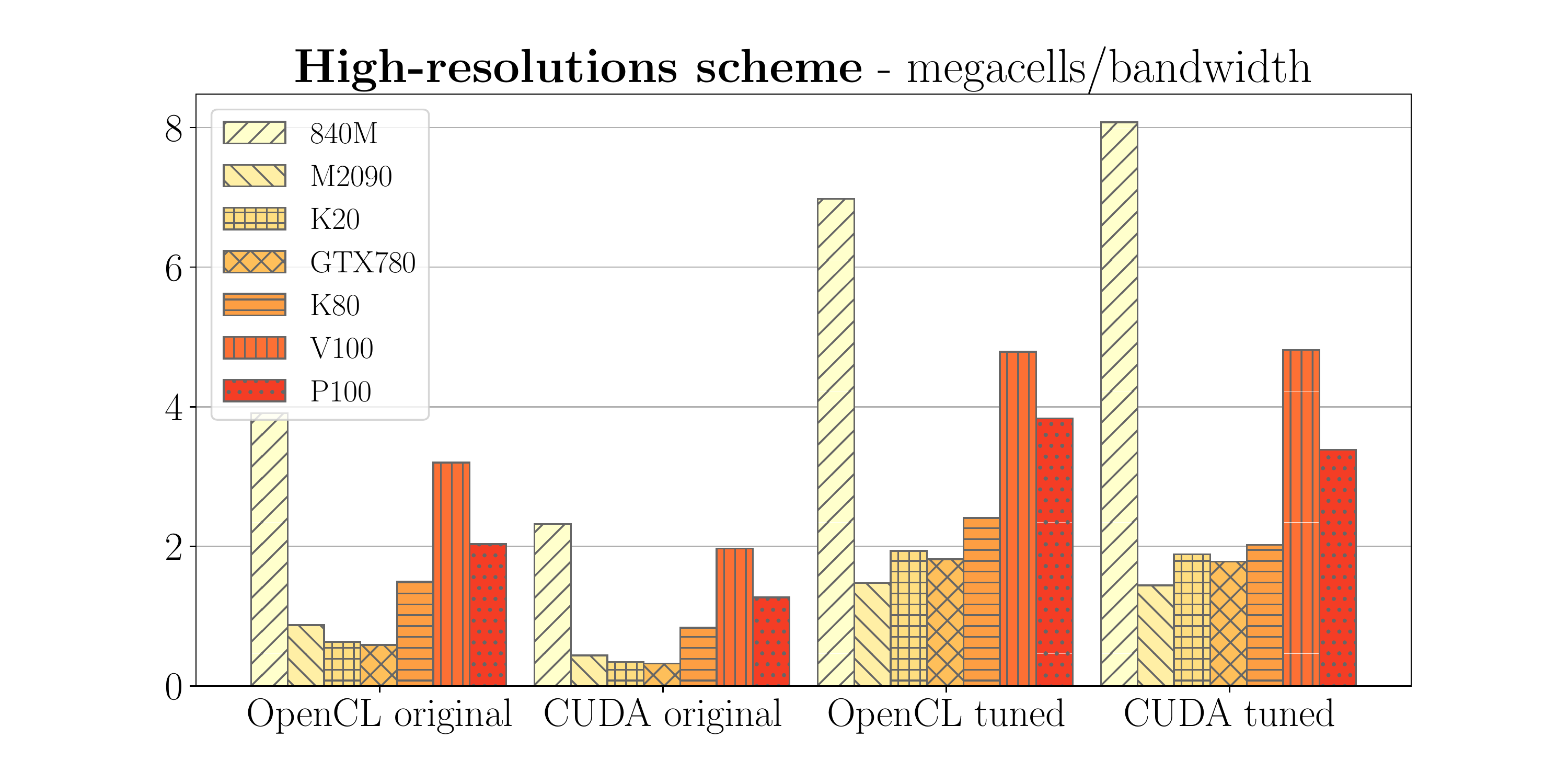}} %
    \hfill\null
    \caption{Performance of original, ported, and optimized kernels measured in megacells per second. The left column is normalized with respect to original performance, and the right column with respect to the theoretical GPU bandwidth. Notice that there is relatively little difference between CUDA and OpenCL, whilst there is a significant difference in how effective the tuning is for the different architectures. Furthermore, there is a significant loss of performance when porting from OpenCL to CUDA in our original approach for the high-resolution and nonlinear schemes. From our experience, this relates to how the two languages optimize mathematical expressions with and without the fast math compilation flags.
    Also notice that the 840M, V100 and P100 GPUs achieve the best performance relative to the GPU bandwidth.} 
    \label{fig:total_results}
    \end{center}
\end{figure*}

The overall performance gain of our optimization is shown in Figure~\ref{fig:total_results}, where all results are given in megacells per second normalized with respect to the original PyOpenCL implementation in the left-hand column and GPU bandwidth (see Table~\ref{tab:hardware}) in the right-hand column. 
The original porting from PyOpenCL to PyCUDA gave a noticeable reduction in performance for the high-resolution scheme on all GPUs. 
After careful examination, we attribute this to different default compilation flags in PyCUDA and PyOpenCL: 
In PyCUDA, the fast-math flag was shown to double the performance for the high-resolution scheme, while we found that it gave less than 5\% performance gain with PyOpenCL. 
Note that the slowdown in the original porting is much less for the linear and nonlinear schemes, as these schemes contain fewer complex mathematical operations, and we instead observe a varying effect on performance of porting the original OpenCL code to CUDA.
When examining the numerical schemes one by one, we see that the optimizations performed for the high-resolution scheme appears to be highly portable when back-ported to PyOpenCL for all GPUs. 
For the tuned nonlinear scheme, however, we see that the 840M and V100 GPUs give significantly higher performance using CUDA than OpenCL.
Finally, for the linear scheme, the performance is similar for all GPUs, and only the 840M and V100 GPUs benefit significantly from the optimization effort. 
In total, we see that certain scheme and GPU combinations result in a significant speedup for CUDA over OpenCL, but we cannot conclude whether this is caused by differences in driver versions or from other factors.
We are therefore not able to claim that CUDA performs better than OpenCL in general.
When looking at the performance normalized with respect to the specific GPU bandwidth, we see that the 840M laptop GPU offers the highest performance for all schemes, followed by the two most recent high-end GPUs, the V100 and P100.

\subsection{Measuring Power Consumption}

We measure power consumption in two ways.
The first method is by using the nvidia-smi application, which can be used to monitor GPU state parameters such as utilization, temperature, power draw, etc.
By programmatically running nvidia-smi in the background during benchmark experiments, we can obtain a log containing a high-resolution power draw profile for the runtime of the benchmark.
The downside of using nvidia-smi is that information about power draw is only supported on recent high-performance GPUs, and we have therefore only benchmarked the power consumption with nvidia-smi on the two most recent Tesla GPUs, the P100 and V100.
Further, nvidia-smi monitors the energy consumption of the GPU only, meaning that we do not have any information about energy consumed by the CPU.
For each experiment, nvidia-smi is started in the background exactly 3~s before the benchmark, and is configured to log the power draw every 20~ms.
This background process is stopped again exactly 3~s after the end of the simulation.
This approach allows us to measure the energy consumption of the idle GPU both before and after each benchmark, and we ignore the idle sections when computing the mean and total power consumption for each experiment.
All results presented here are with the idle load subtracted from the experiment results.

The second method is to measure the total amount of energy used by the entire computer through a watt meter.
The use of the watt meter requires physical access to the computer, and we are therefore restricted to do measurements on the laptop and desktop, containing the GeForce 840M and GeForce GTX780 GPUs, respectively.
The watt meter offers no automatic logging or reading, but displays the total power used with an accuracy of 1 Wh.
To get sufficiently accurate readings we need to run each benchmark long enough to keep the GPU busy for approximately one hour, after which we read the total and mean power consumption for each experiment.
Before and after each benchmark, we also record the background power of the idle system, and the maximum recorded power during the experiment, to monitor whether the operating system is putting any non-related background load on the computer.
It should also be noted that the battery was removed from the laptop during these experiments.
Similarly to the first method, we subtract the idle loads from the result of each experiment, but note that the addition increased load from the CPU will still be included.

\subsection{Comparing Energy Efficiency}

\newcommand{\figw}{0.318}

\begin{figure}
	\begin{center}
    \hfill
    	\includegraphics[width=\figw\linewidth, trim=1.0cm 0.0cm 1.7cm 0.0cm, clip]{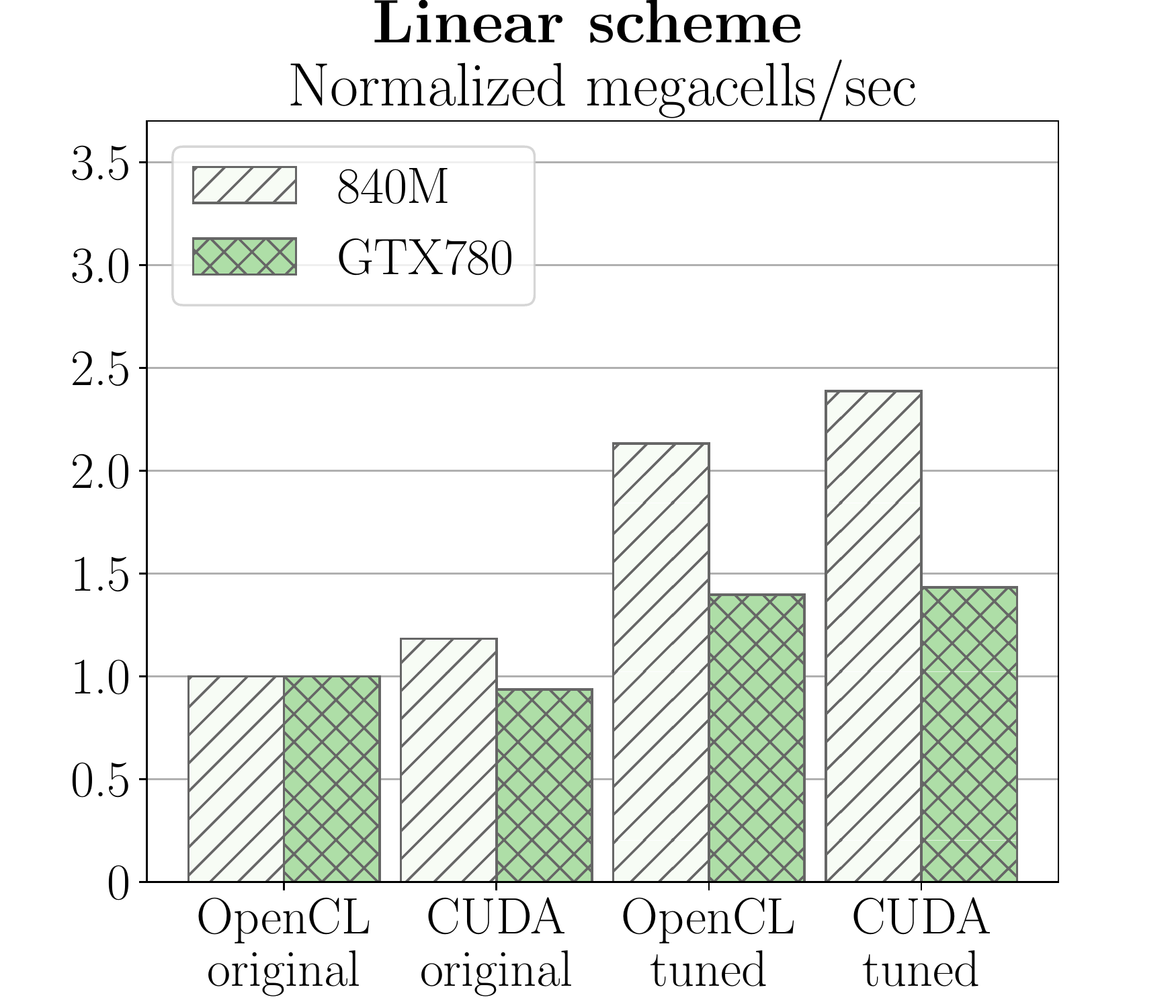}
    	\,
    	\includegraphics[width=\figw\linewidth, trim=1.0cm 0.0cm 1.7cm 0.0cm, clip]{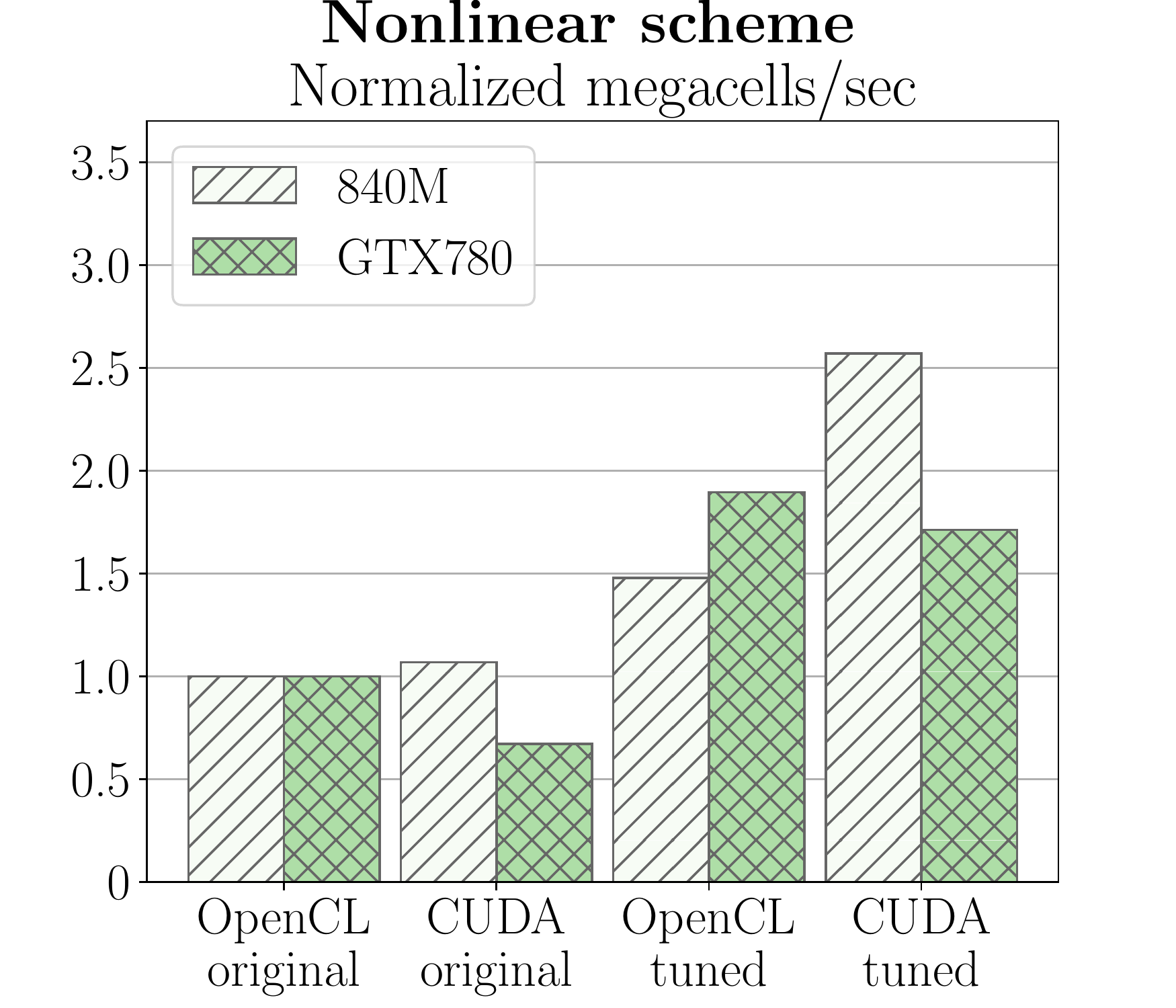}
        \,
    	\includegraphics[width=\figw\linewidth, trim=1.0cm 0.0cm 1.7cm 0.0cm, clip]{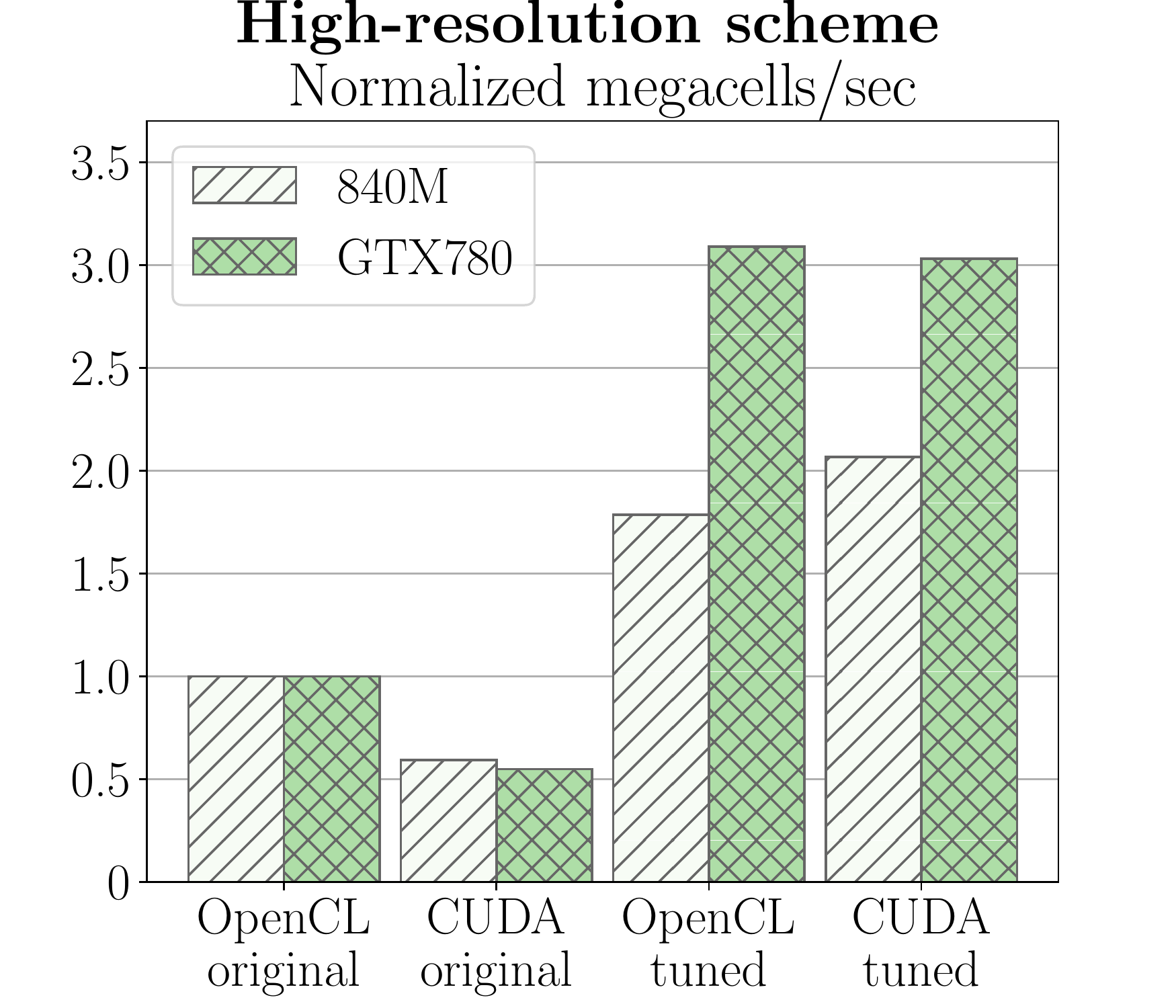}
    \hfill\null
    \\
    \hfill
    	\includegraphics[width=\figw\linewidth, trim=1.0cm 0.0cm 1.7cm 0.9cm, clip]{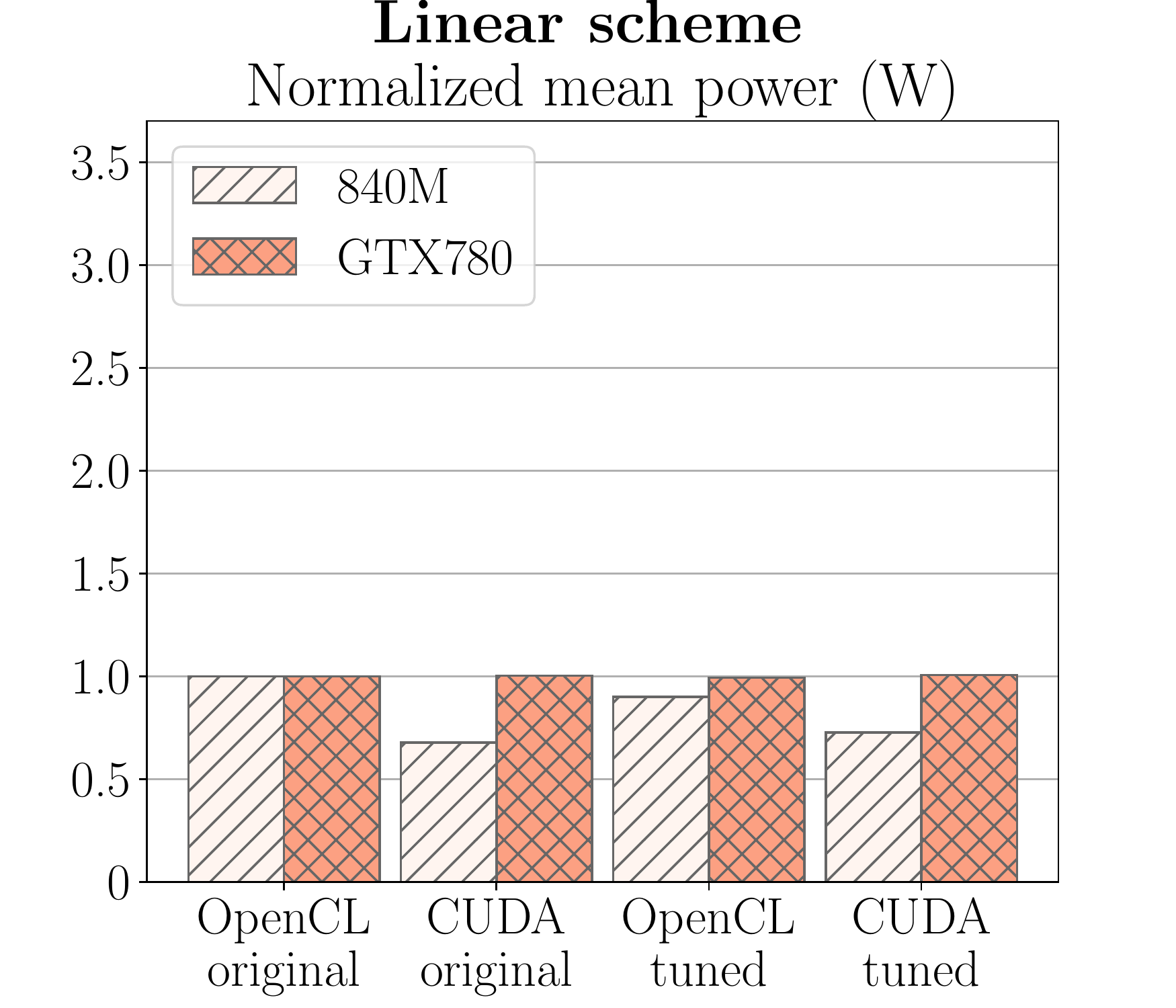}
    	\,
    	\includegraphics[width=\figw\linewidth, trim=1.0cm 0.0cm 1.7cm 0.9cm, clip]{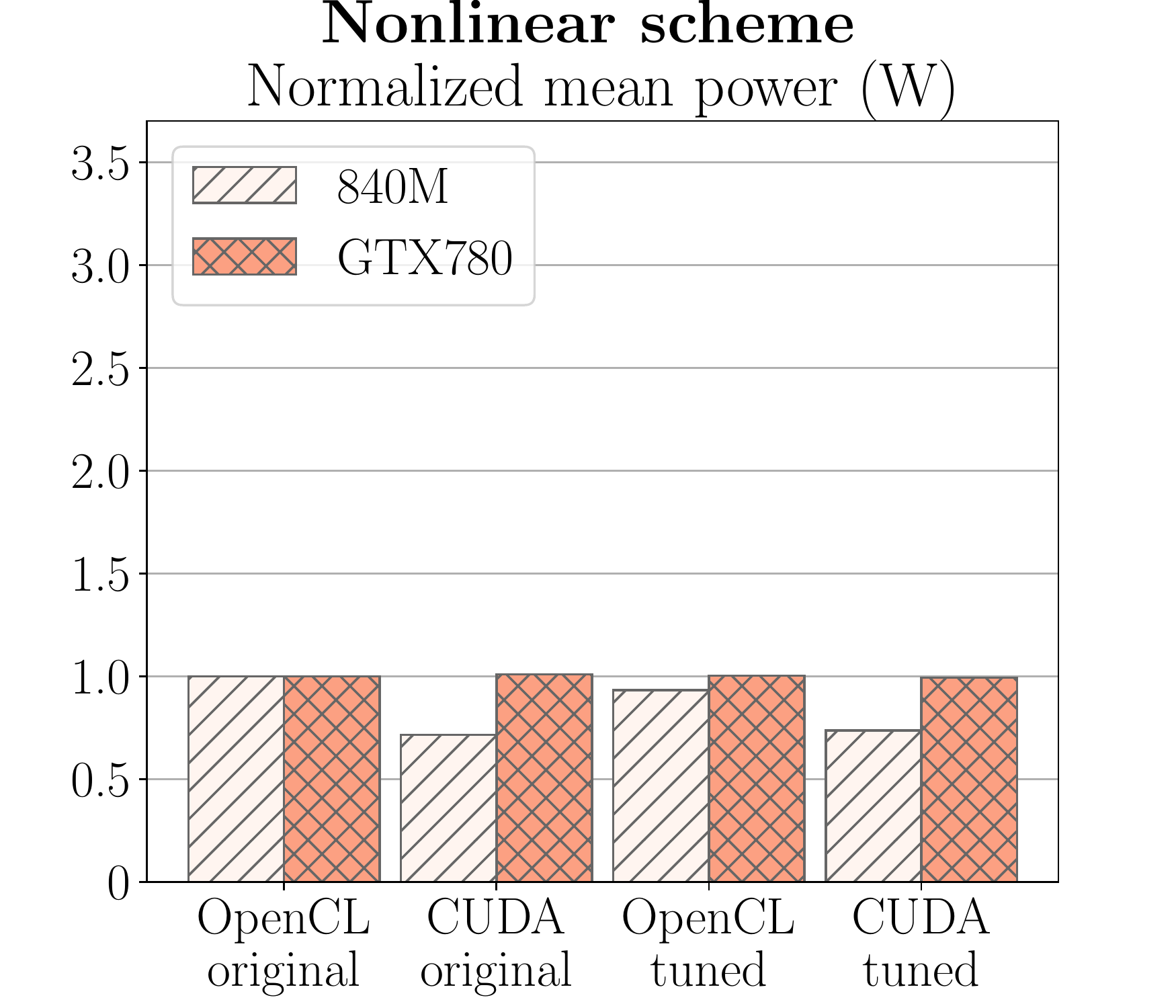}
        \,
    	\includegraphics[width=\figw\linewidth, trim=1.0cm 0.0cm 1.7cm 0.9cm, clip]{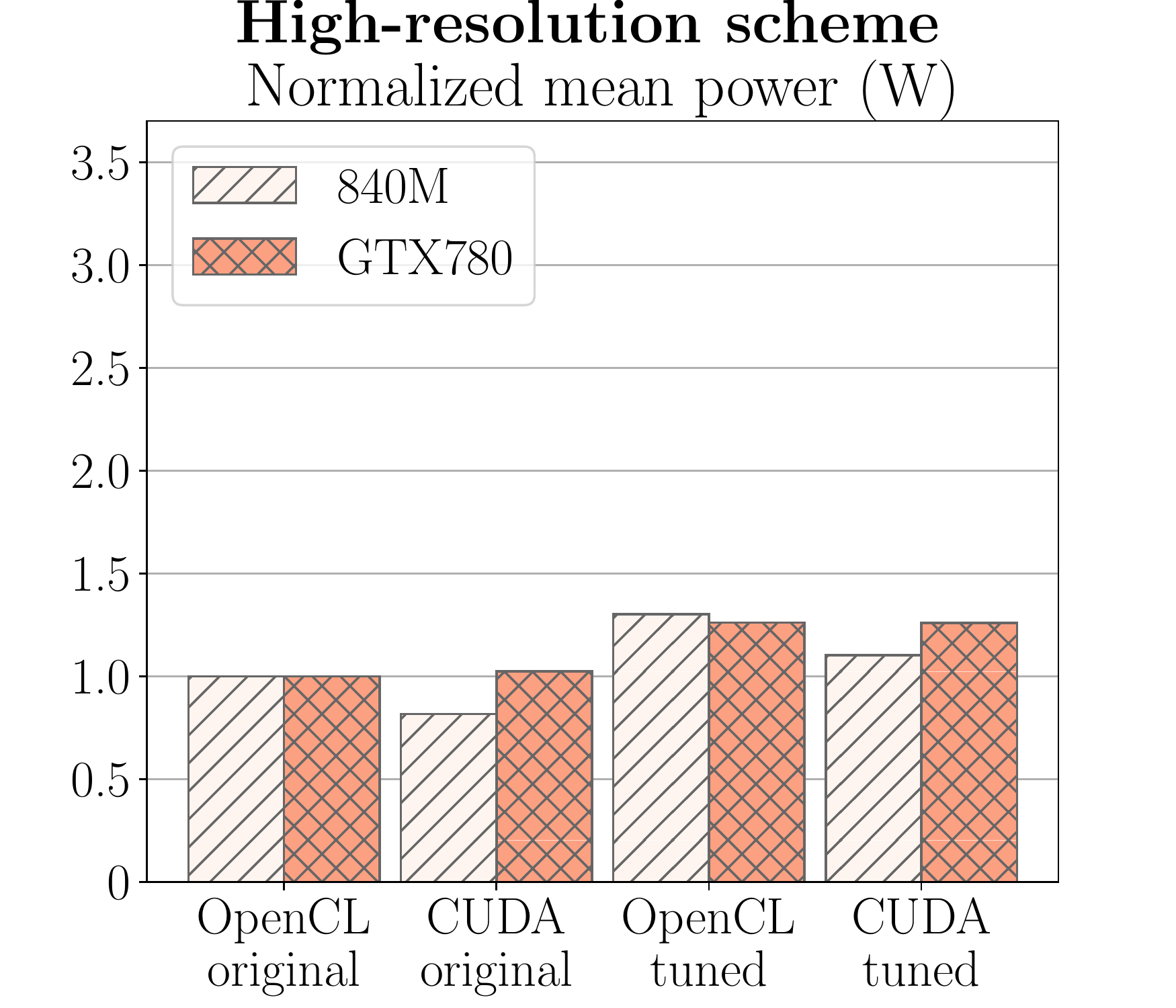}
    \hfill\null
    \\
    \hfill
    	\includegraphics[width=\figw\linewidth, trim=1.0cm 0.0cm 1.7cm 0.9cm, clip]{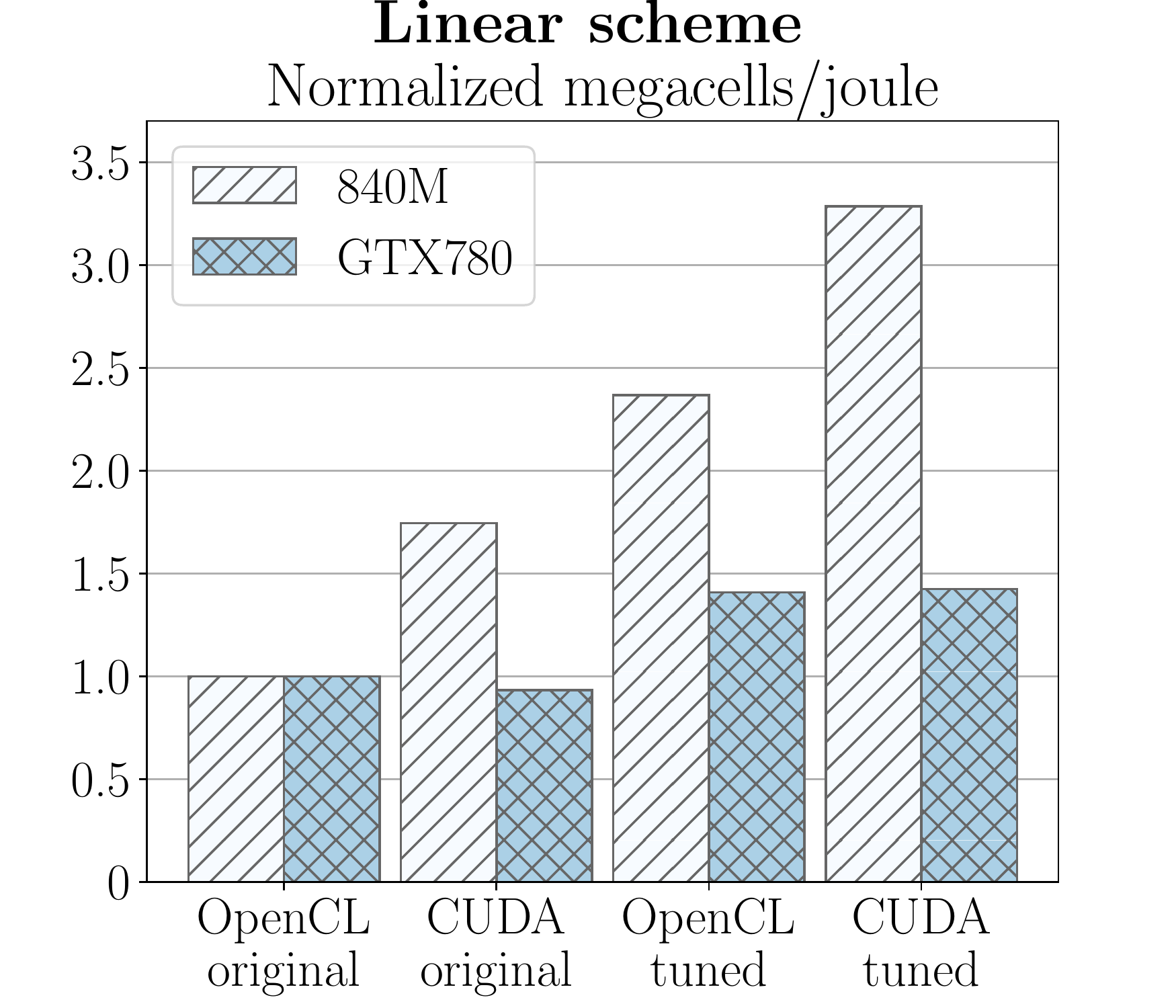}
    	\,
    	\includegraphics[width=\figw\linewidth, trim=1.0cm 0.0cm 1.7cm 0.9cm, clip]{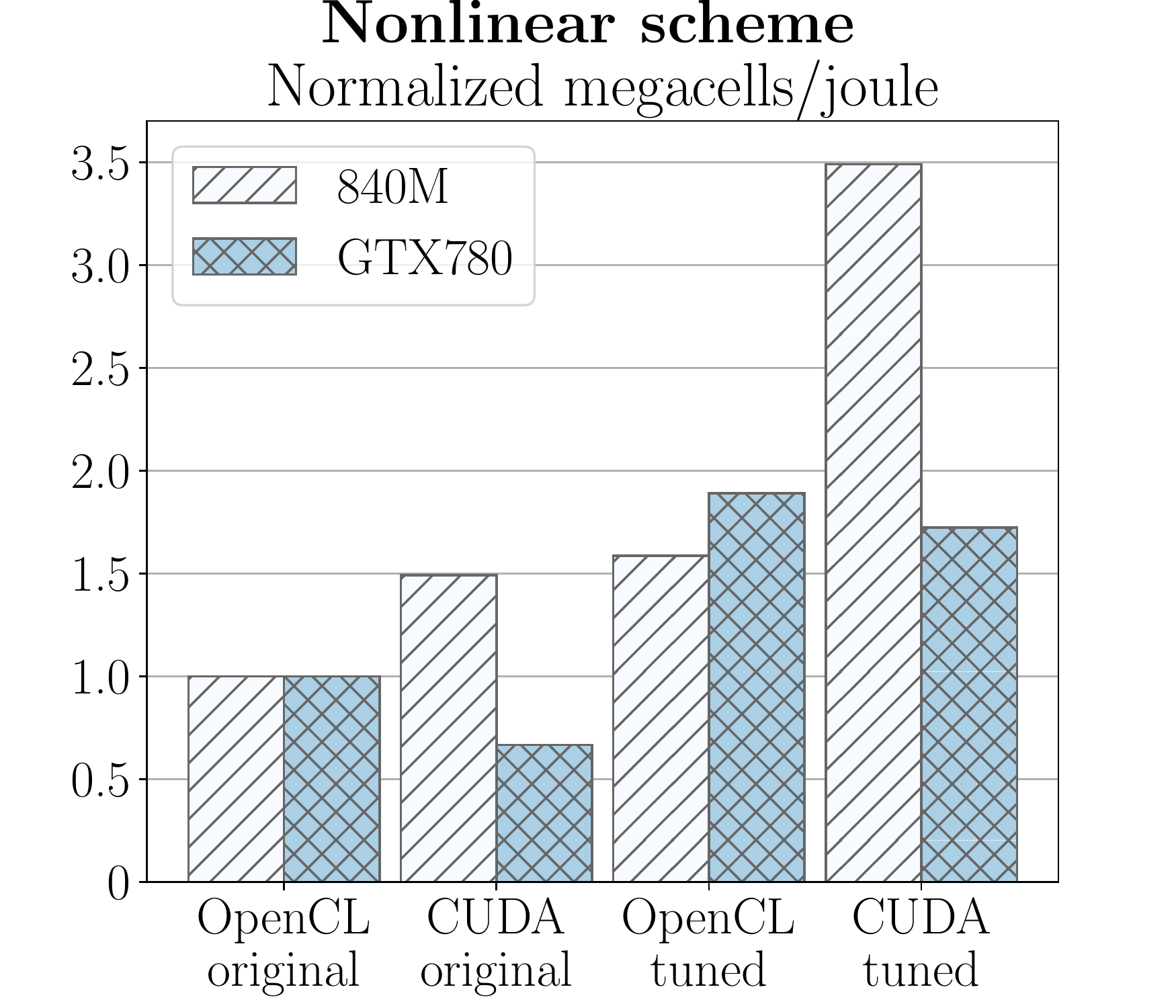}
        \,
    	\includegraphics[width=\figw\linewidth, trim=1.0cm 0.0cm 1.7cm 0.9cm, clip]{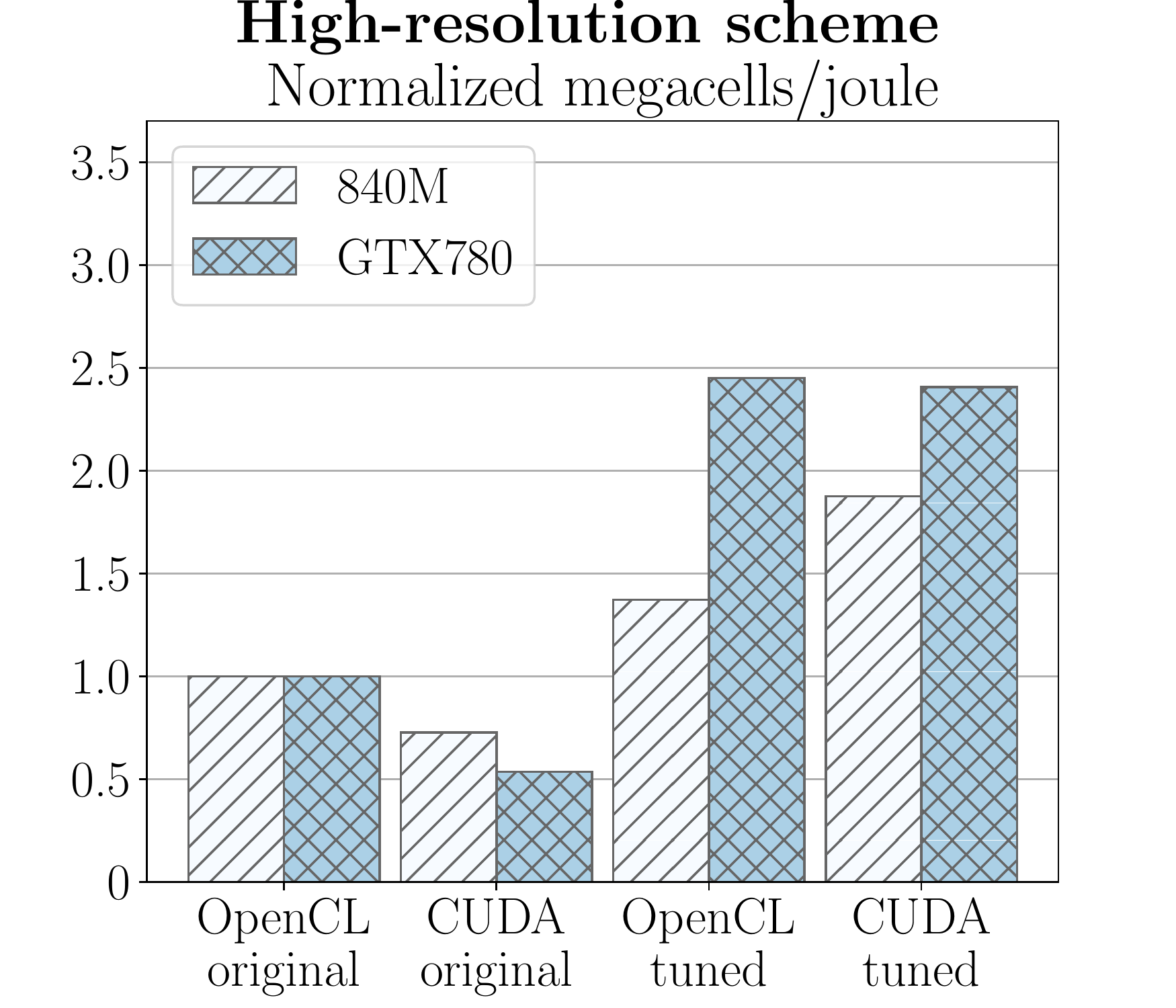}
    \hfill\null
    \caption{
    	Comparison of original, ported and optimized codes measured in megacells per second (top row), mean power usage (mid row), and megacells per joule (bottom row), normalized with respect to the original OpenCL implementation, for the laptop (840M) and desktop (GTX780) GPUs.
    	The power is measured through a watt meter, and represents the power consumed by the entire computer.
    	Note that the CUDA versions requires less power than the OpenCL versions on the 840M, whereas there are no differences between equivalent versions on the GTX780.
    	In terms of power efficiency, CUDA is more efficient than OpenCL on the M840, whereas the GTX780 gives the same power efficiency.}
    \label{fig:power_portability_watt_meter}
    \end{center}
\end{figure}

\begin{figure}
	\begin{center}
    \hfill
    	\includegraphics[width=\figw\linewidth, trim=1.0cm 0.0cm 1.7cm 0.0cm, clip]{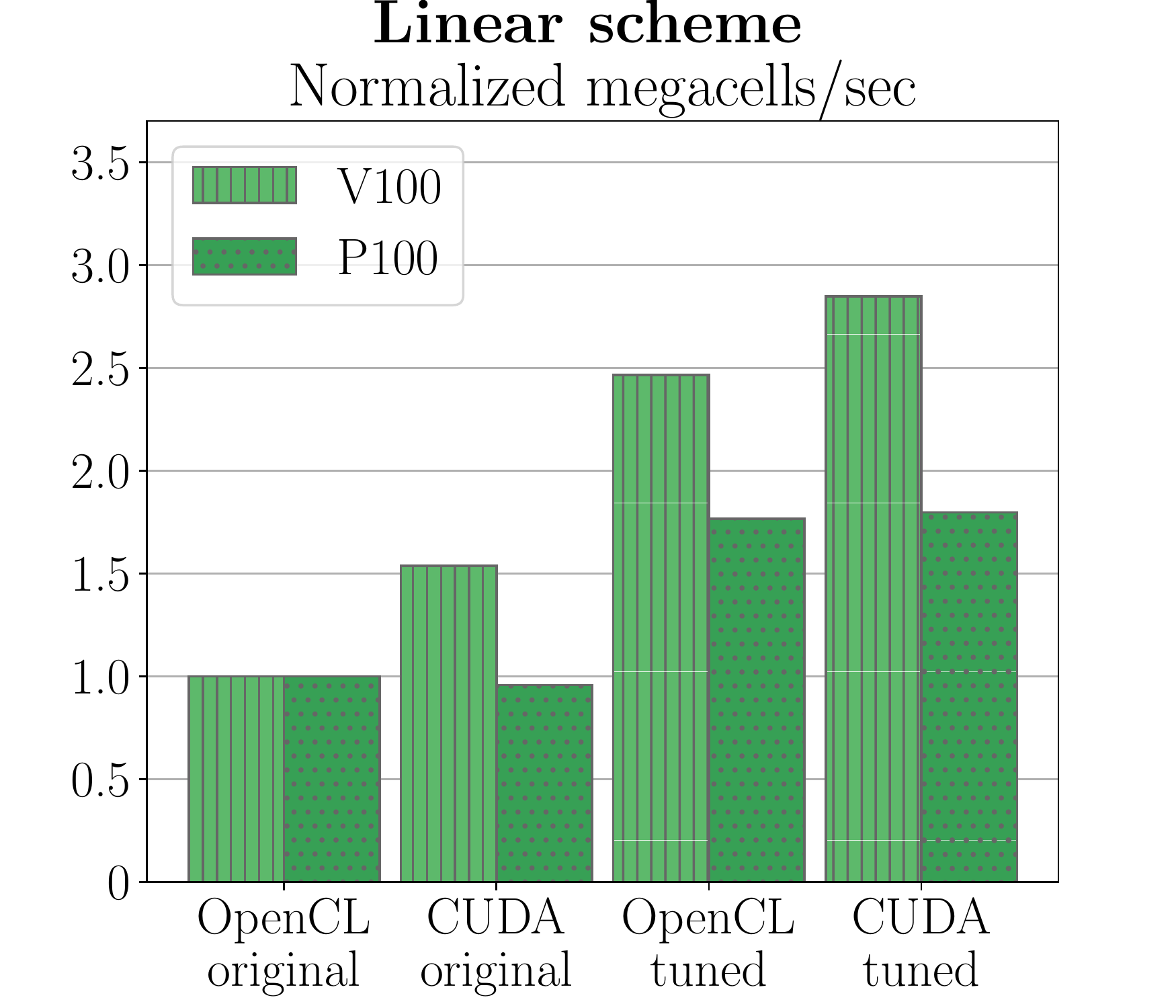}
    	\,
    	\includegraphics[width=\figw\linewidth, trim=1.0cm 0.0cm 1.7cm 0.0cm, clip]{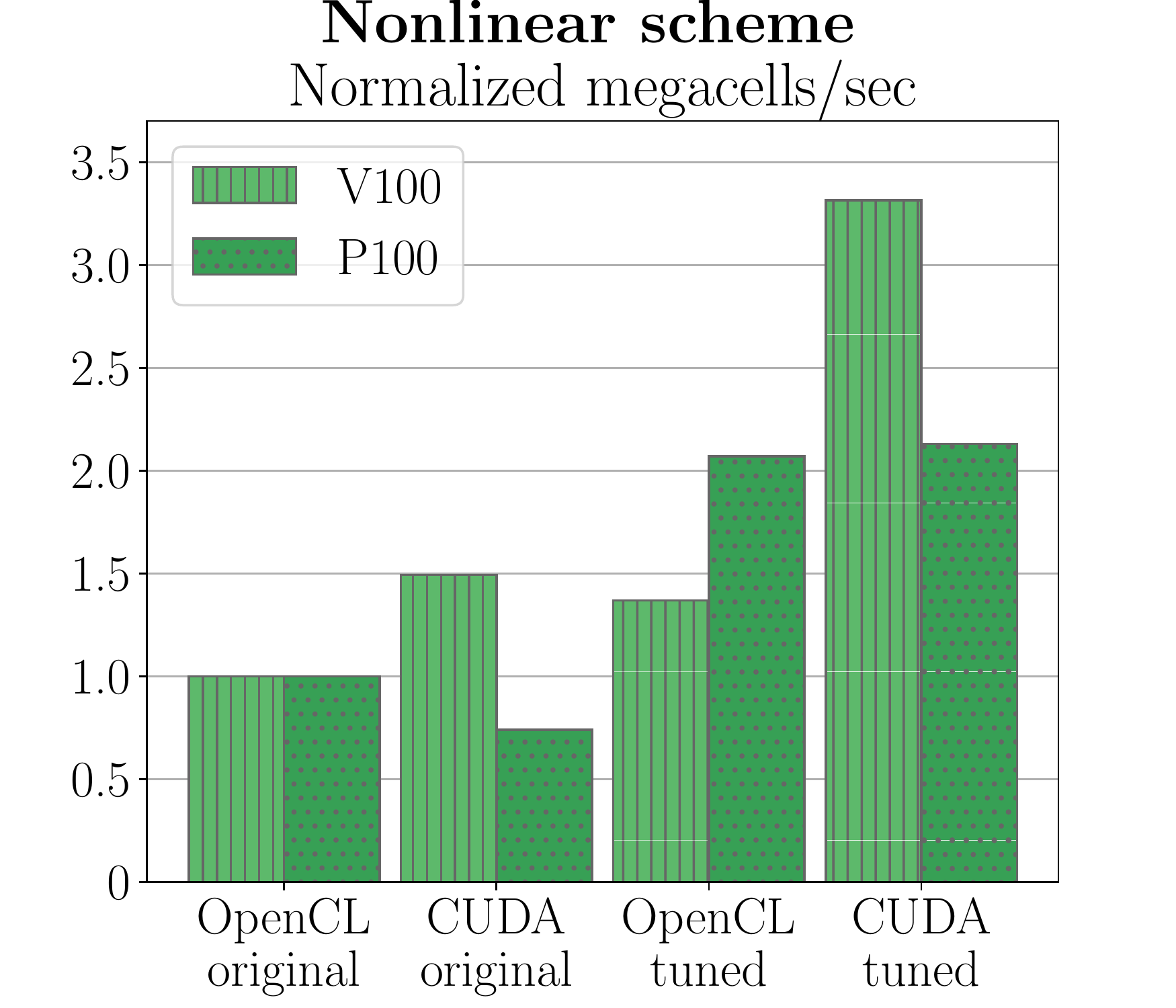}
        \,
    	\includegraphics[width=\figw\linewidth, trim=1.0cm 0.0cm 1.7cm 0.0cm, clip]{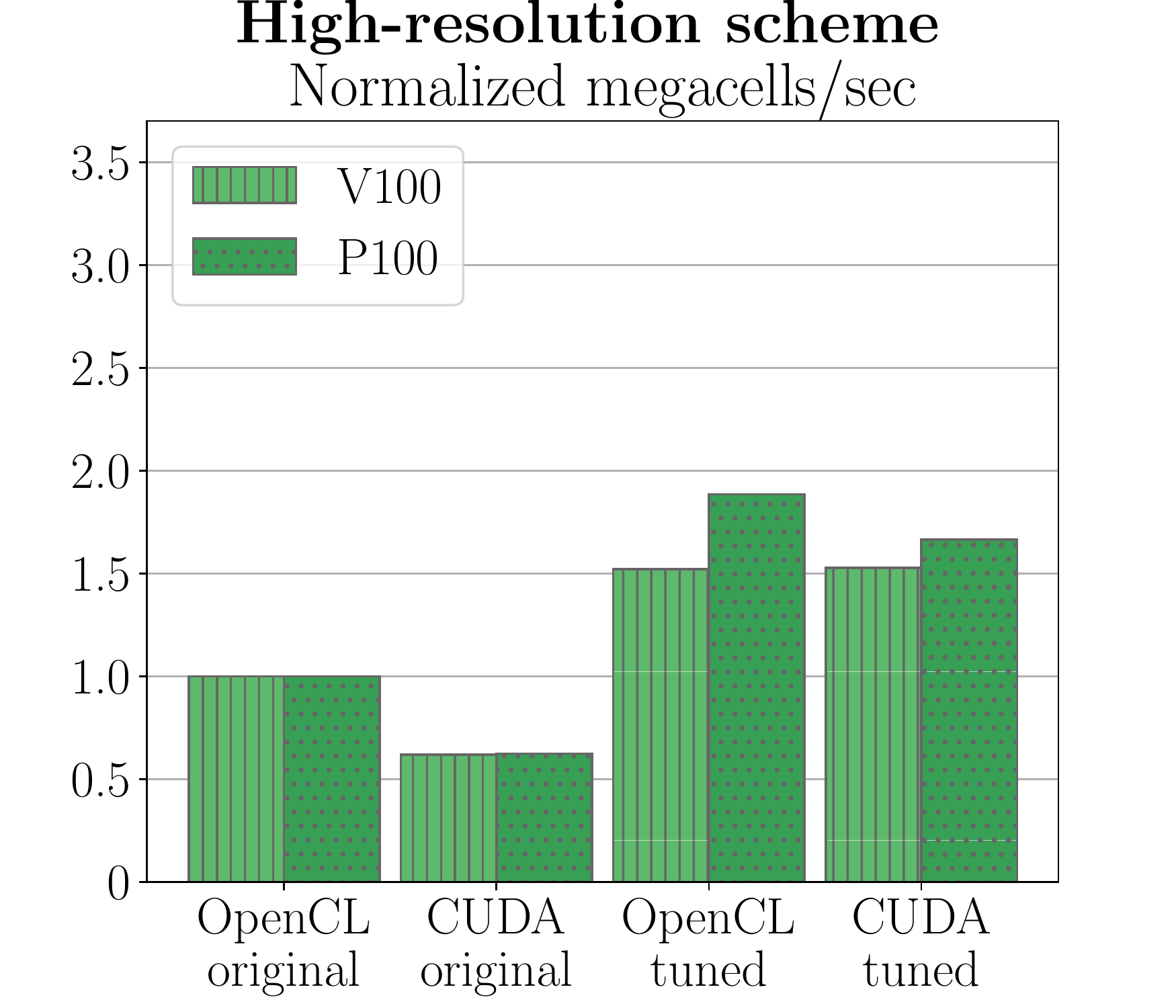}
    \hfill\null
    \\
    \hfill
    	\includegraphics[width=\figw\linewidth, trim=1.0cm 0.0cm 1.7cm 0.9cm, clip]{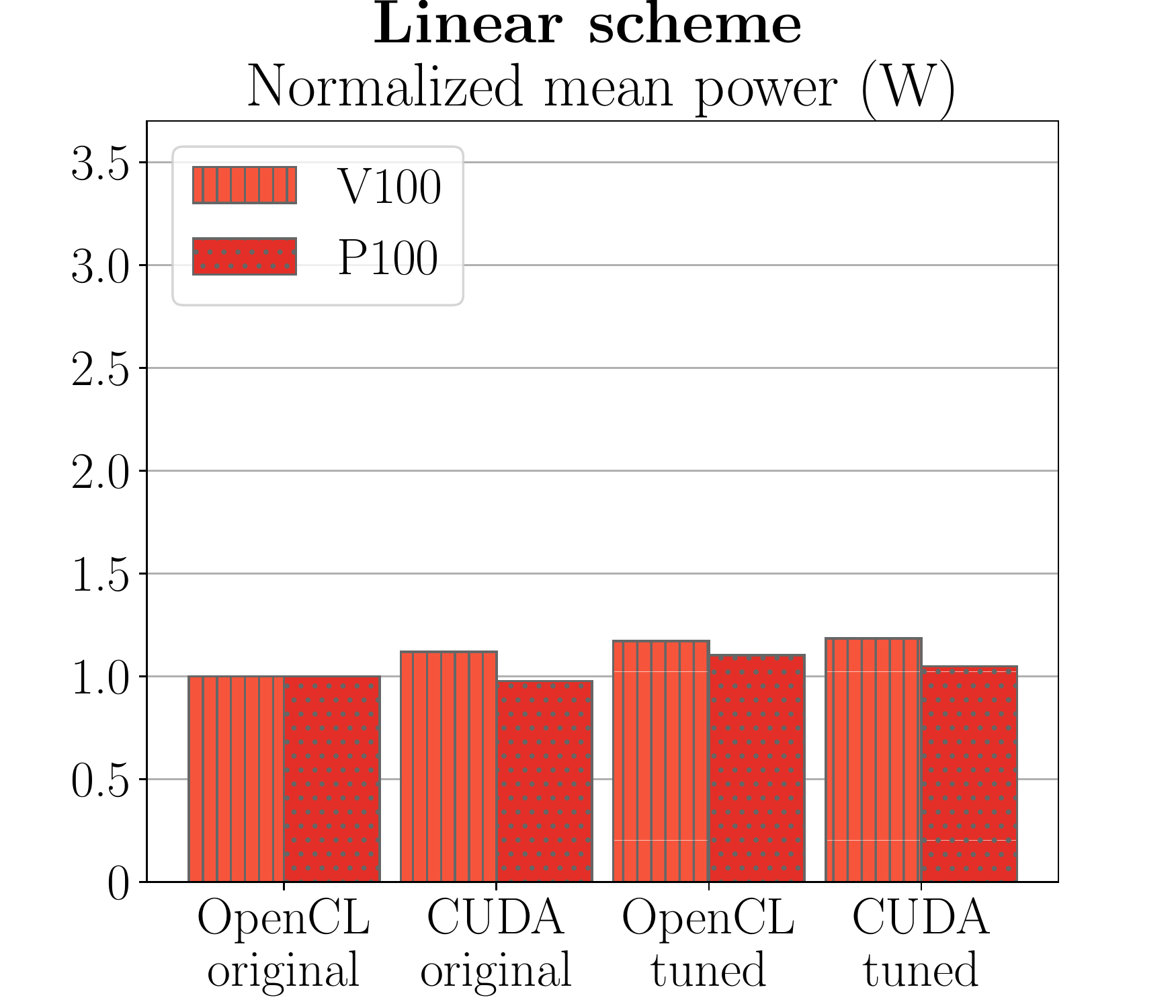}
    	\,
    	\includegraphics[width=\figw\linewidth, trim=1.0cm 0.0cm 1.7cm 0.9cm, clip]{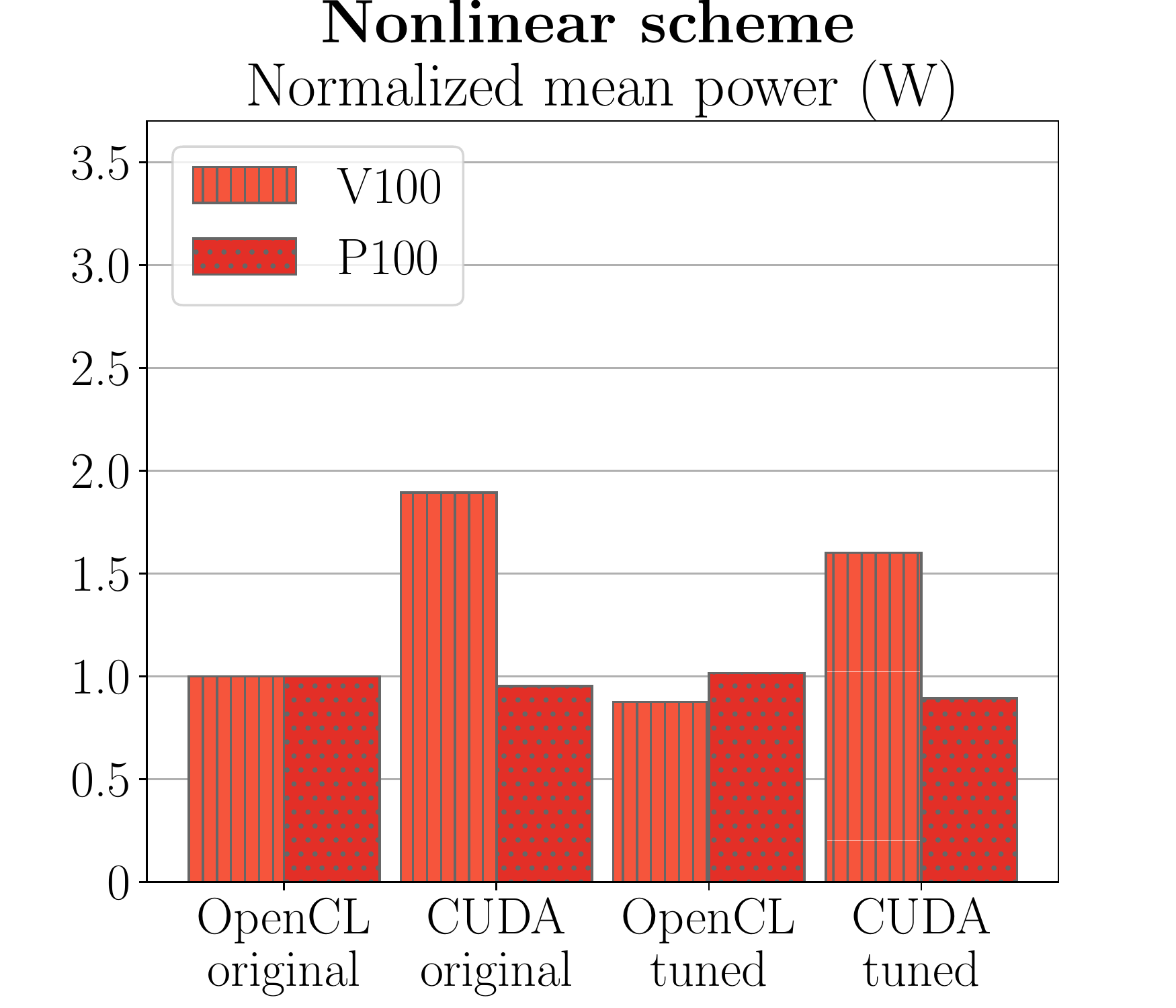}
        \,
    	\includegraphics[width=\figw\linewidth, trim=1.0cm 0.0cm 1.7cm 0.9cm, clip]{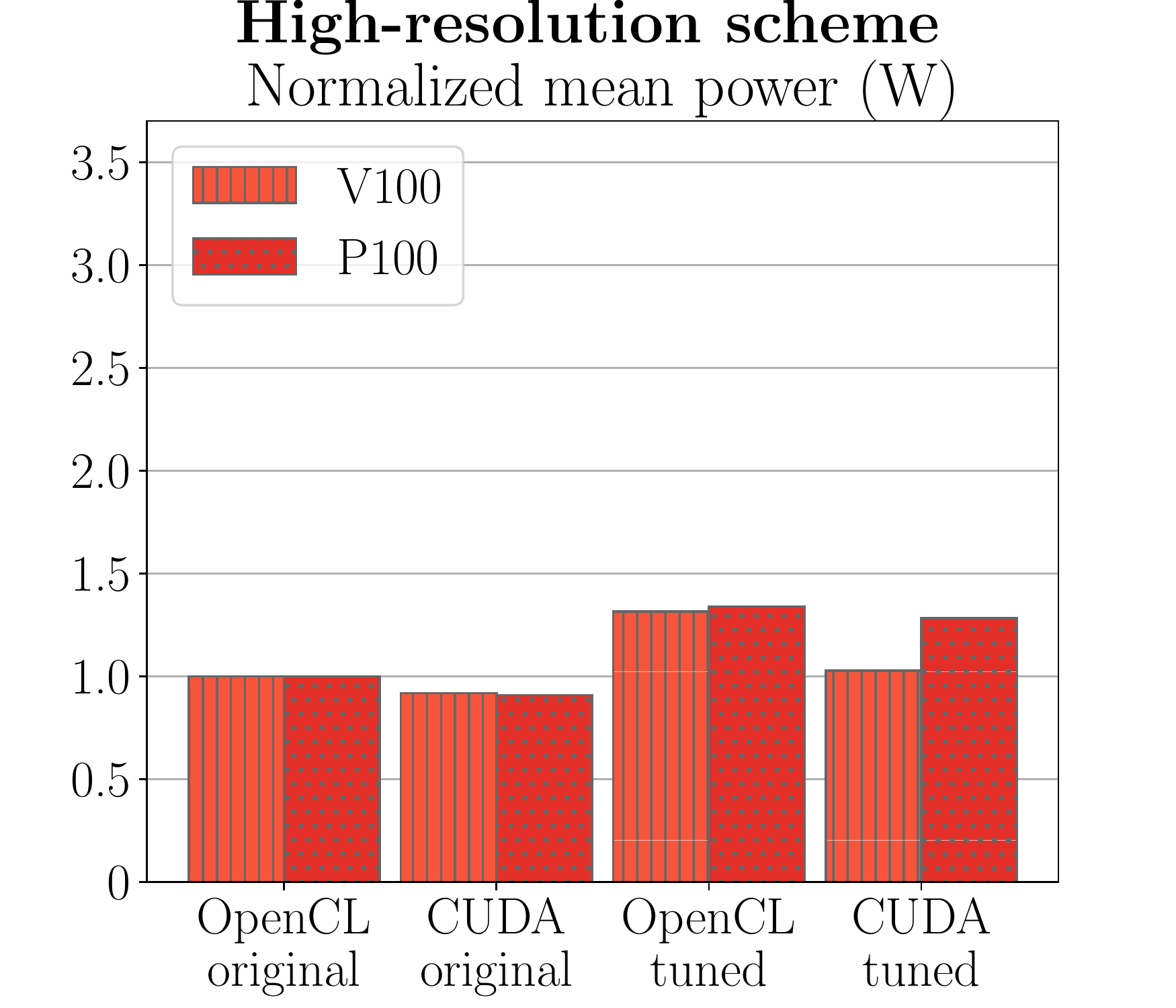}
    \hfill\null
    \\
    \hfill
    	\includegraphics[width=\figw\linewidth, trim=1.0cm 0.0cm 1.7cm 0.9cm, clip]{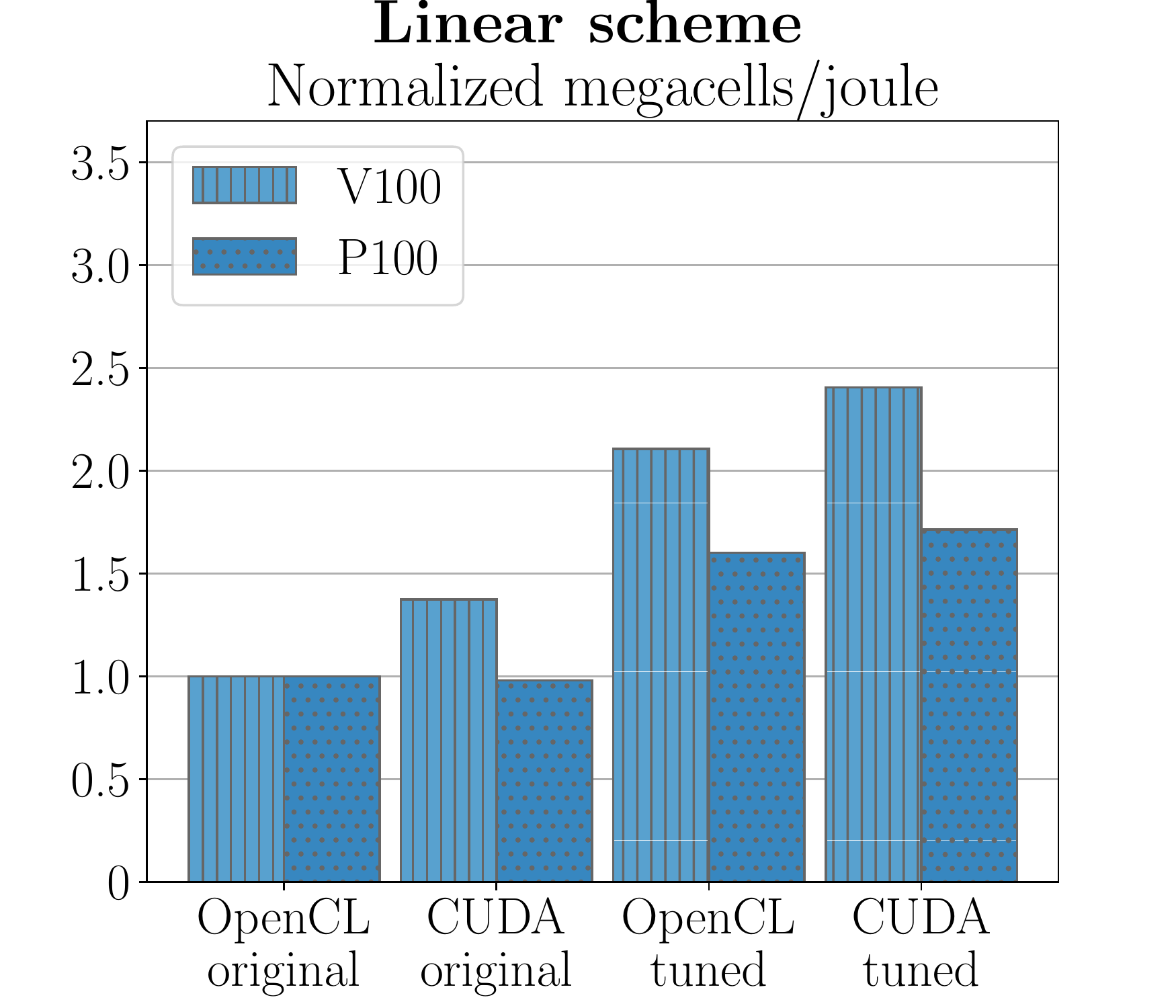}
    	\,
    	\includegraphics[width=\figw\linewidth, trim=1.0cm 0.0cm 1.7cm 0.9cm, clip]{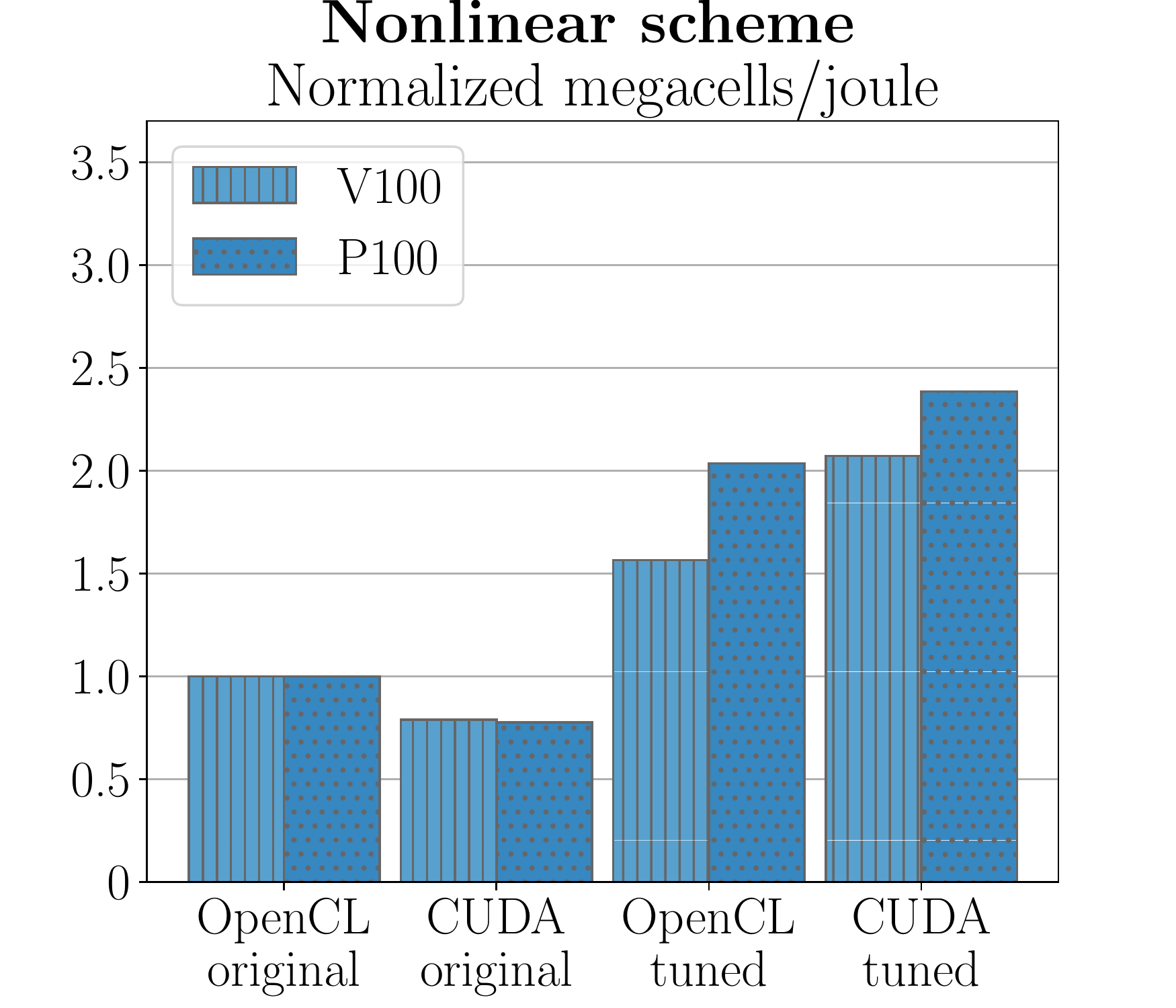}
        \,
    	\includegraphics[width=\figw\linewidth, trim=1.0cm 0.0cm 1.7cm 0.9cm, clip]{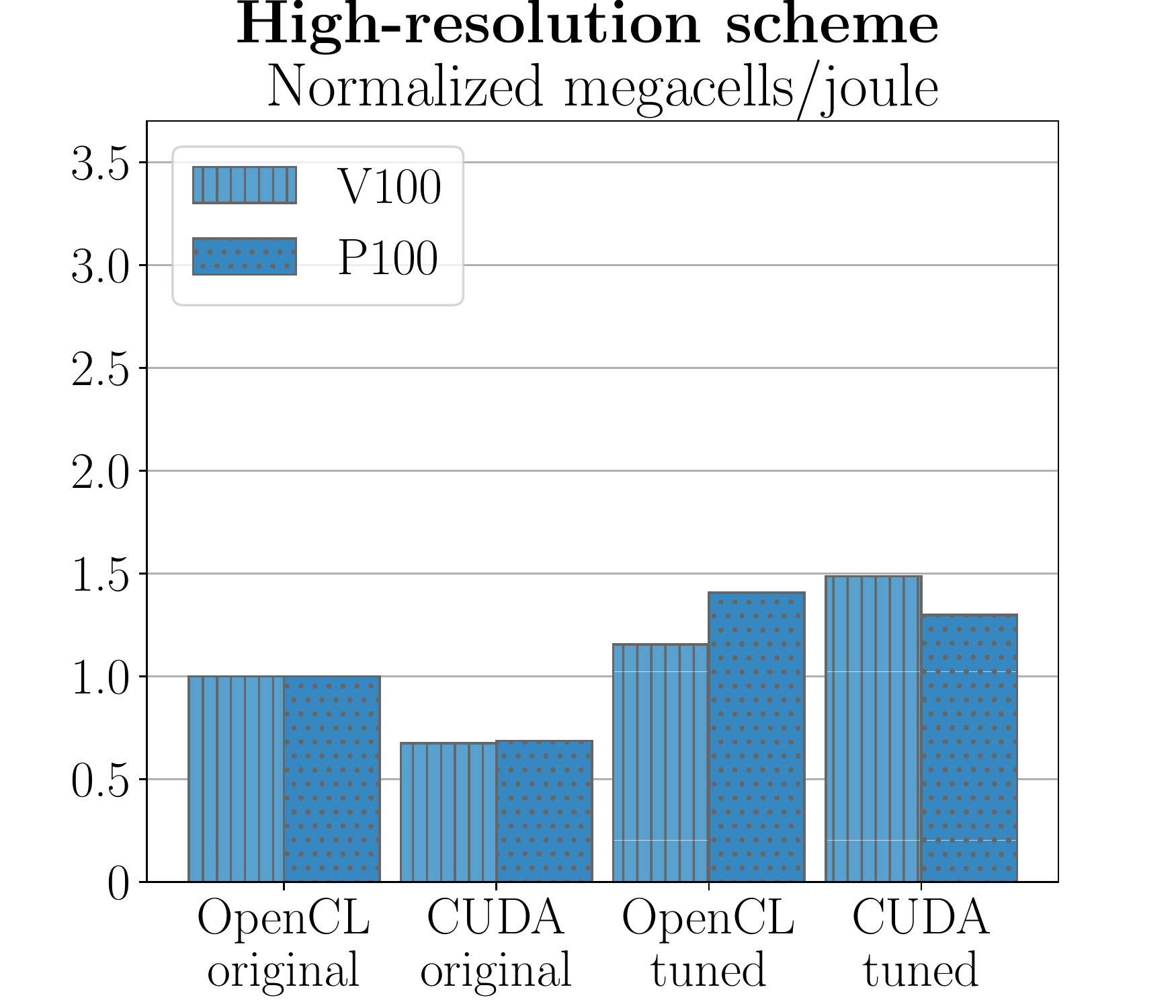}
    \hfill\null
    \caption{
        Comparison of original, ported and optimized codes measured in megacells per second (top row), mean power usage (mid row), and megacells per joule (bottom row), normalized with respect to the original OpenCL implementation, for the Tesla P100 and V100 GPUs.
        The power is measured through nvidia-smi, and represents the power consumed by the GPU only.
    	On the V100, there are only minor differences in mean power consumption between different versions of the linear and high-resolution scheme, but CUDA uses more power than OpenCL for the nonlinear scheme.
    	CUDA is, however, more power efficient than OpenCL for all three tuned schemes on this GPU.
    	On the P100, the mean power consumption is equal for all versions of the linear and nonlinear schemes, which means that power efficiency becomes equivalent to computational efficiency. For the high-resolution scheme we see that the mean power consumption increases for the tuned codes, but the computational performance increases more, meaning that the power efficiency is still improved by 30--40\%.}
    \label{fig:power_portability_nvidia_smi}
    \end{center}
\end{figure}

Figure~\ref{fig:power_portability_watt_meter} shows the results from the power efficiency experiments using the watt meter on the laptop (840M) and desktop (GTX780).
The top row repeats the results for computational performance also shown in Figure~\ref{fig:total_results} for the relevant GPUs, whereas the second row show the normalized mean power consumption with respect to the original OpenCL versions.
The first thing we notice is that CUDA seems to require less power on the 840M compared to OpenCL for all versions of the three schemes.
On the GTX780, however, there are no differences between the two programming models for equivalent versions. 
In fact, only the tuned high-resolution scheme seems to be different from the others (using about 30\% more power), and this behavior can also be seen on the 840M.
The power efficiency of the three schemes is shown in the bottom row in Figure~\ref{fig:power_portability_watt_meter}, and we see that on the 840M the tuned CUDA versions are the most power efficient.
This is because CUDA is both more efficient and uses less power on this system.
On the GTX780, CUDA and OpenCL have equivalent power efficiency for all tuned schemes.

The results for the Tesla P100 and Tesla V100 are shown in Figure~\ref{fig:power_portability_nvidia_smi}.
The top row shows the computational performance in megacells per second, repeating the results from Figure~\ref{fig:total_results}.
The second row shows the mean power used by each version of the three schemes.
Note here that on the V100, both CUDA versions for the nonlinear scheme use 60--90\% more power than the OpenCL versions, which is the opposite result compared to the 840M results.
For both the linear and high-resolution schemes, the results do not differ significantly in favor of either CUDA or OpenCL, but the tuned OpenCL version uses slightly more power for the high-resolution scheme.
When we consider the power efficiency in the bottom row, we see that the tuned CUDA versions are the best versions for all schemes.
In particular for the nonlinear scheme, this is mostly due to the large difference in computational efficiency between CUDA and OpenCL for this particular scheme on this particular GPU.
For the P100, however, the mean power consumption is almost the same for all versions of each scheme, except for the high-resolution scheme which has an increased mean power consumption for the tuned versions.
The increase in mean power consumption is however less than the increase in computational performance, and the high-resolution scheme also has a higher energy efficiency when tuned.

In general, we observe that all experiments show a mean power usage within about 30\% of the original OpenCL versions, with the exception of the nonlinear scheme on the V100.
On the other side, the computational performance increases up to 5 times (the high-resolution scheme on the GTX780).
This shows that the most important factor for improving power efficiency is to increase computational performance.

\subsection{Tuning Block Size Configuration for Energy Efficiency}
In the previous section, all benchmarks were configured using the optimal block size configuration for computational efficiency.
Here, we analyze how sensitive power efficiency is to the block size configuration, and whether there is other optimal configurations if we aim for power efficiency instead of computational performance.

  \newcommand{\figwbs}{0.28}
 \begin{figure}
	\begin{center}
    \hfill
    	\includegraphics[width=\figwbs\linewidth, trim=0.6cm 2.6cm 1.2cm 1.8cm, clip]{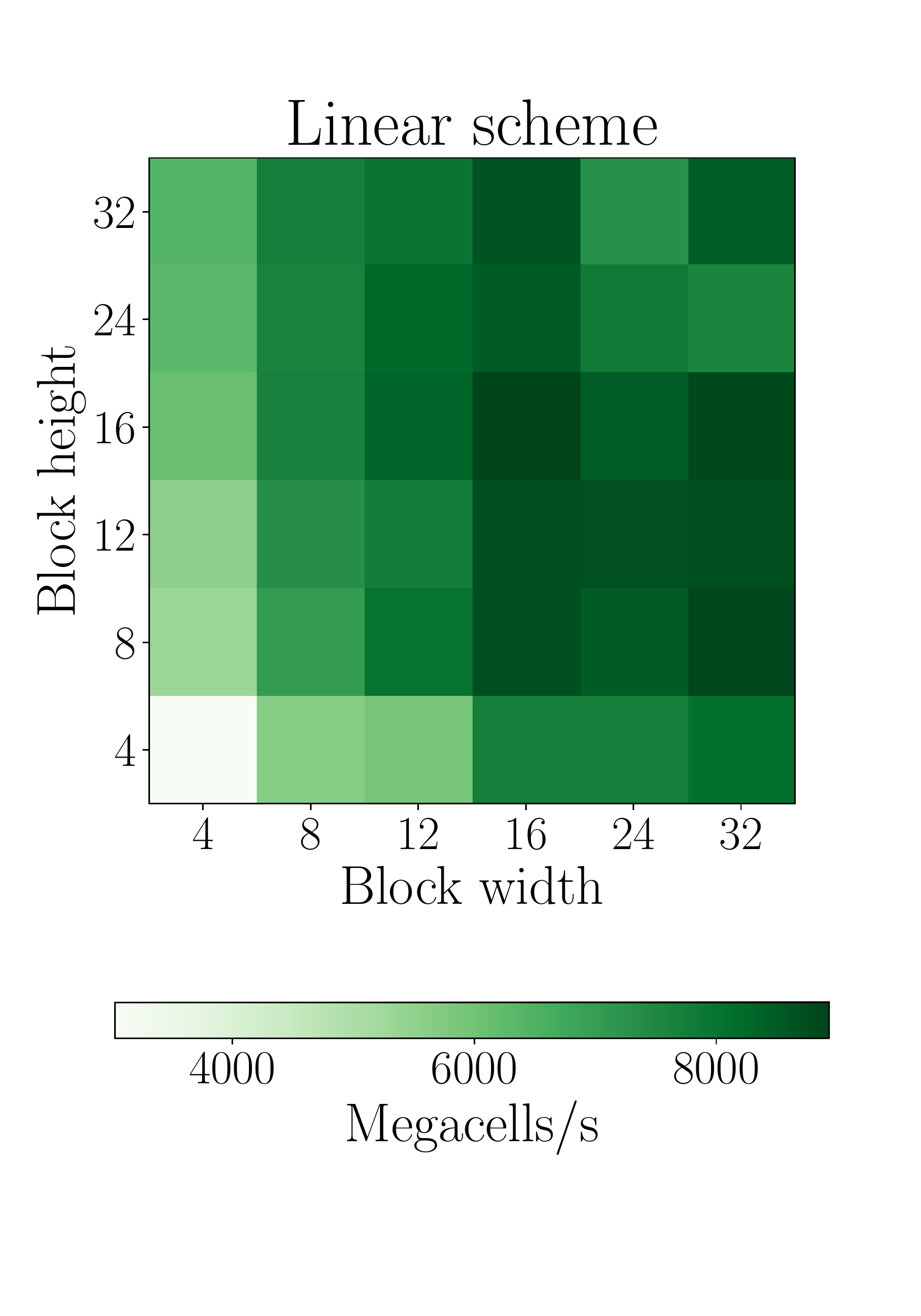}
    	\,
    	\includegraphics[width=\figwbs\linewidth, trim=0.6cm 2.6cm 1.2cm 1.8cm, clip]{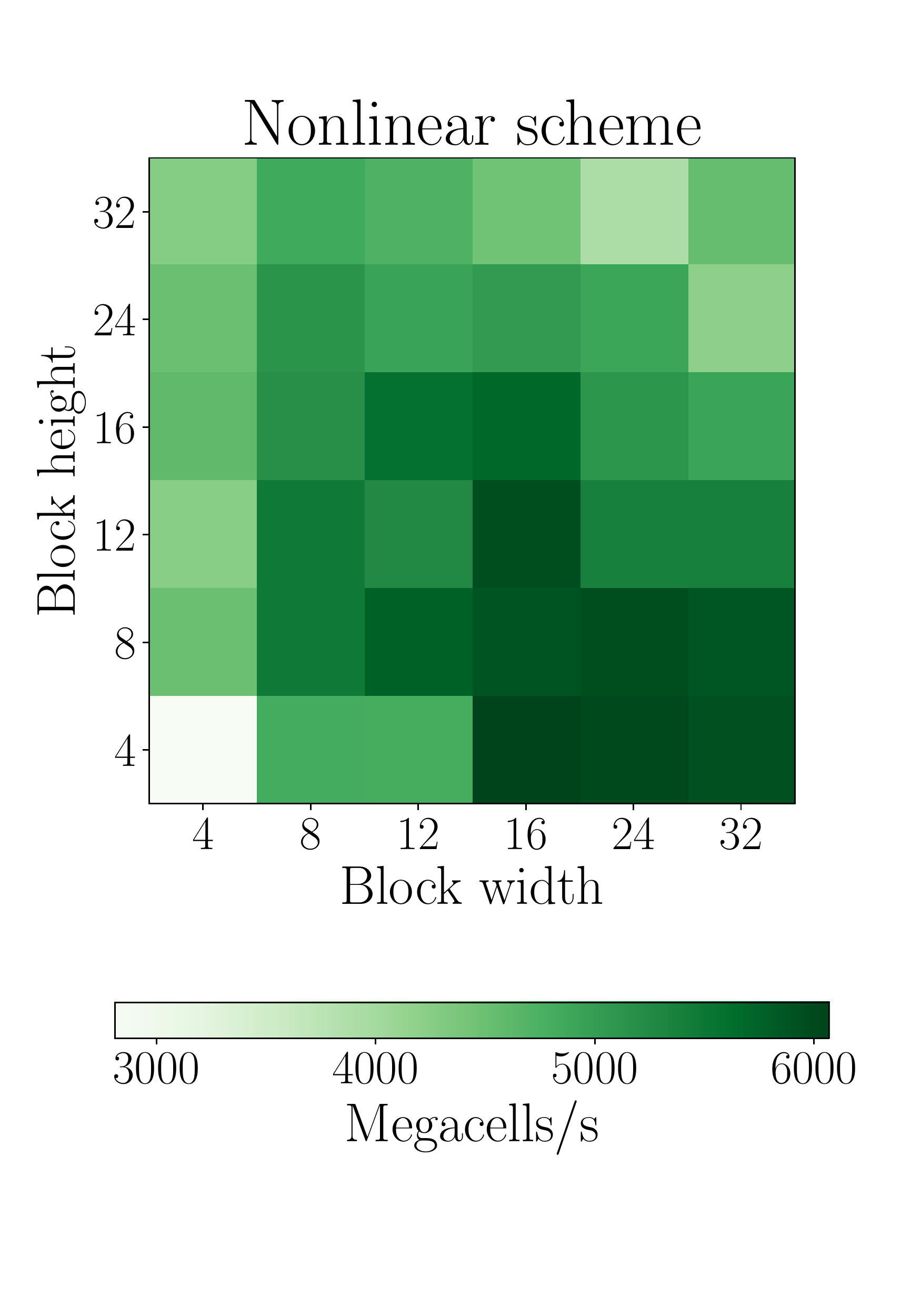}
        \,
    	\includegraphics[width=\figwbs\linewidth, trim=0.6cm 2.6cm 1.2cm 1.8cm, clip]{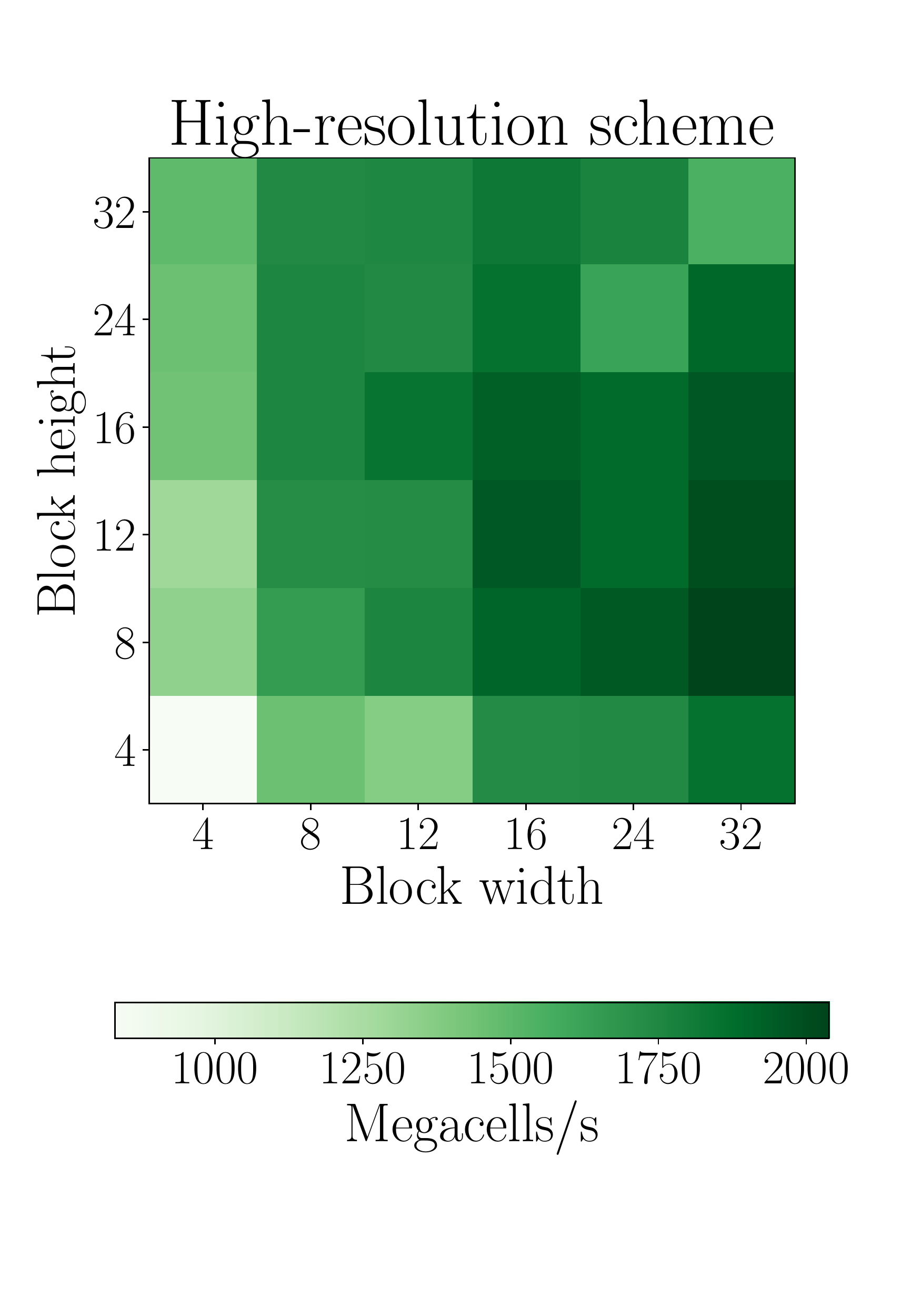}
    \hfill\null
    \\ 
    \hfill
    	\includegraphics[width=\figwbs\linewidth, trim=0.6cm 2.6cm 1.2cm 3.0cm, clip]{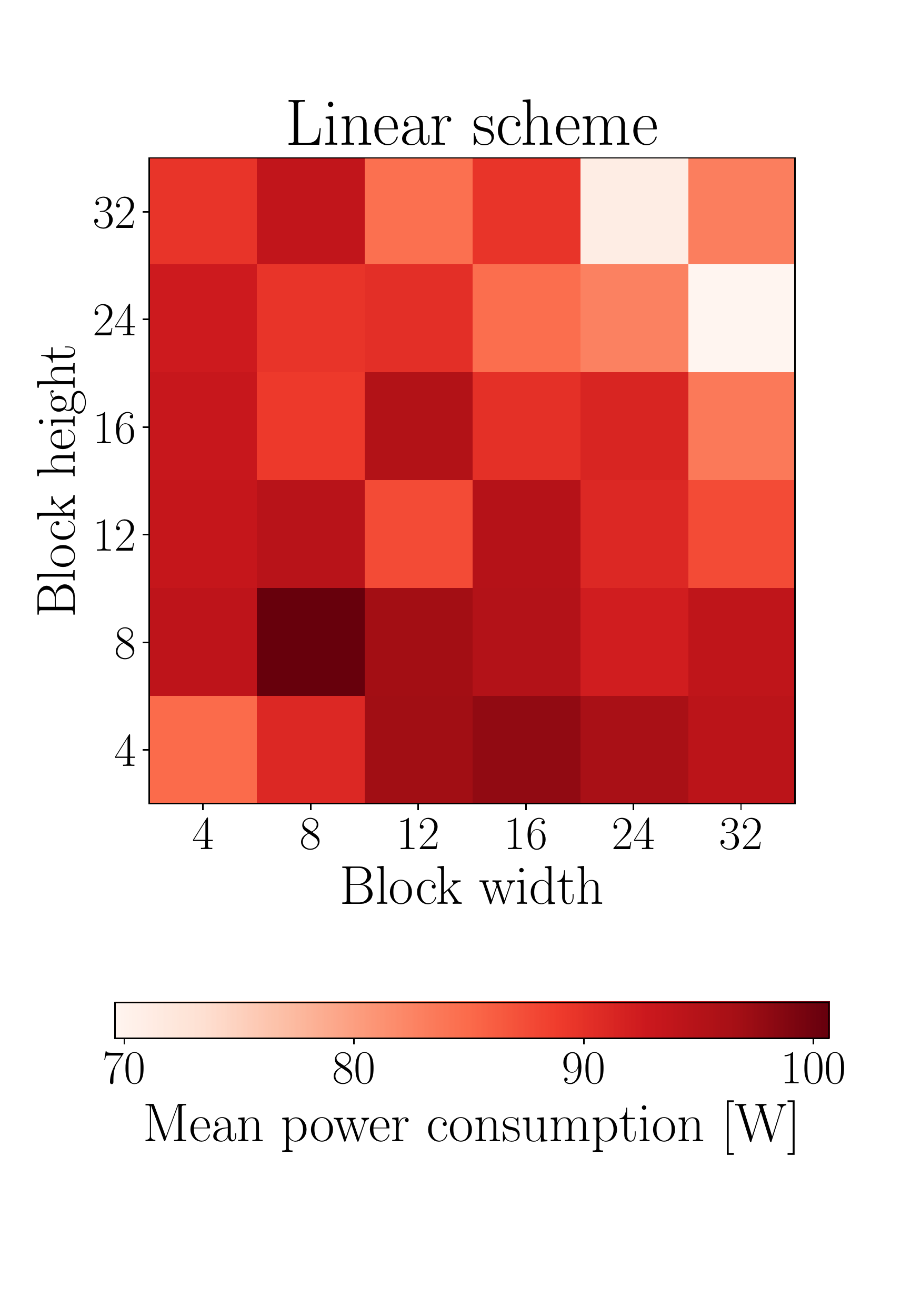}
    	\,
    	\includegraphics[width=\figwbs\linewidth, trim=0.6cm 2.6cm 1.2cm 3.0cm, clip]{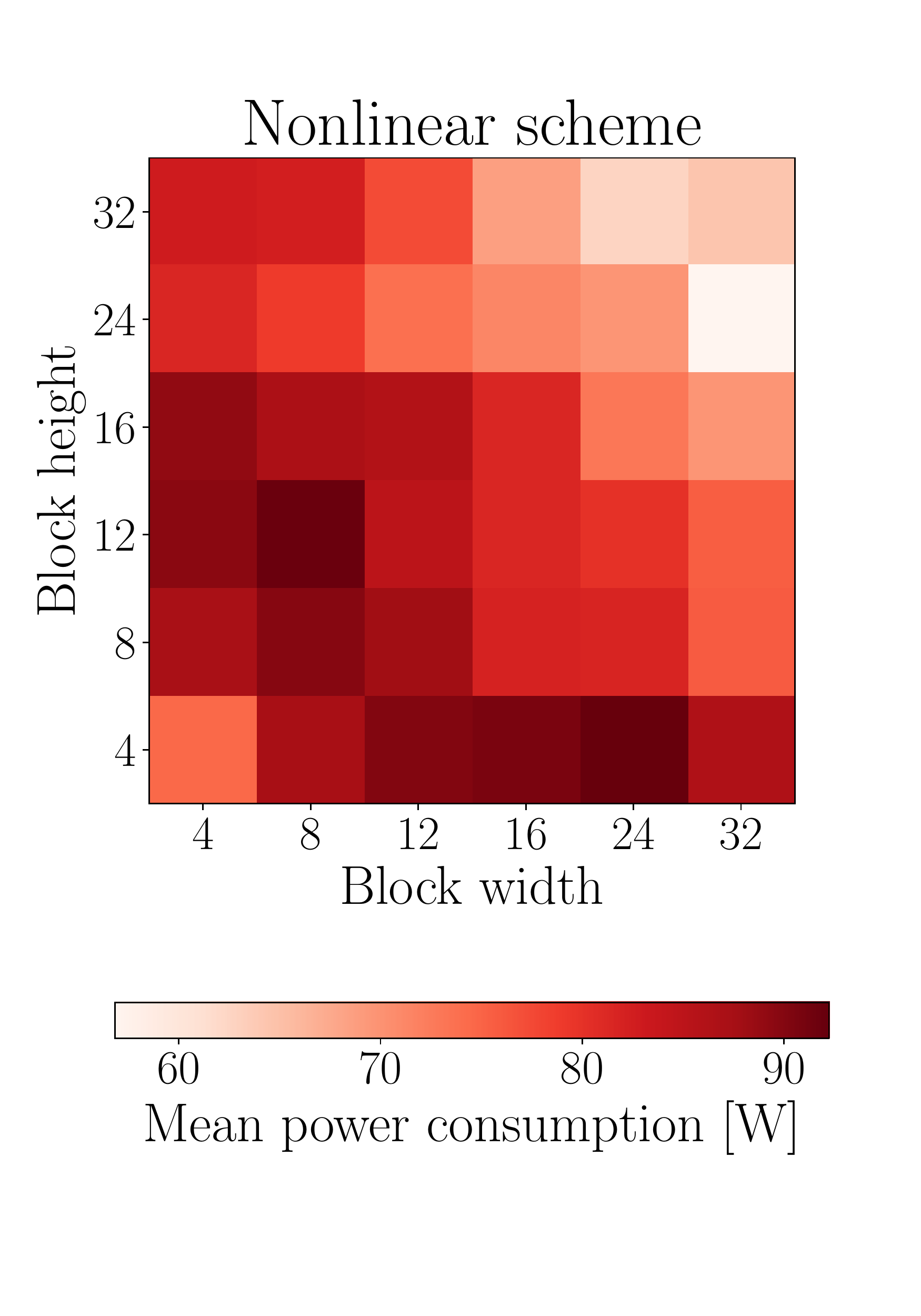}
        \,
    	\includegraphics[width=\figwbs\linewidth, trim=0.6cm 2.6cm 1.2cm 3.0cm, clip]{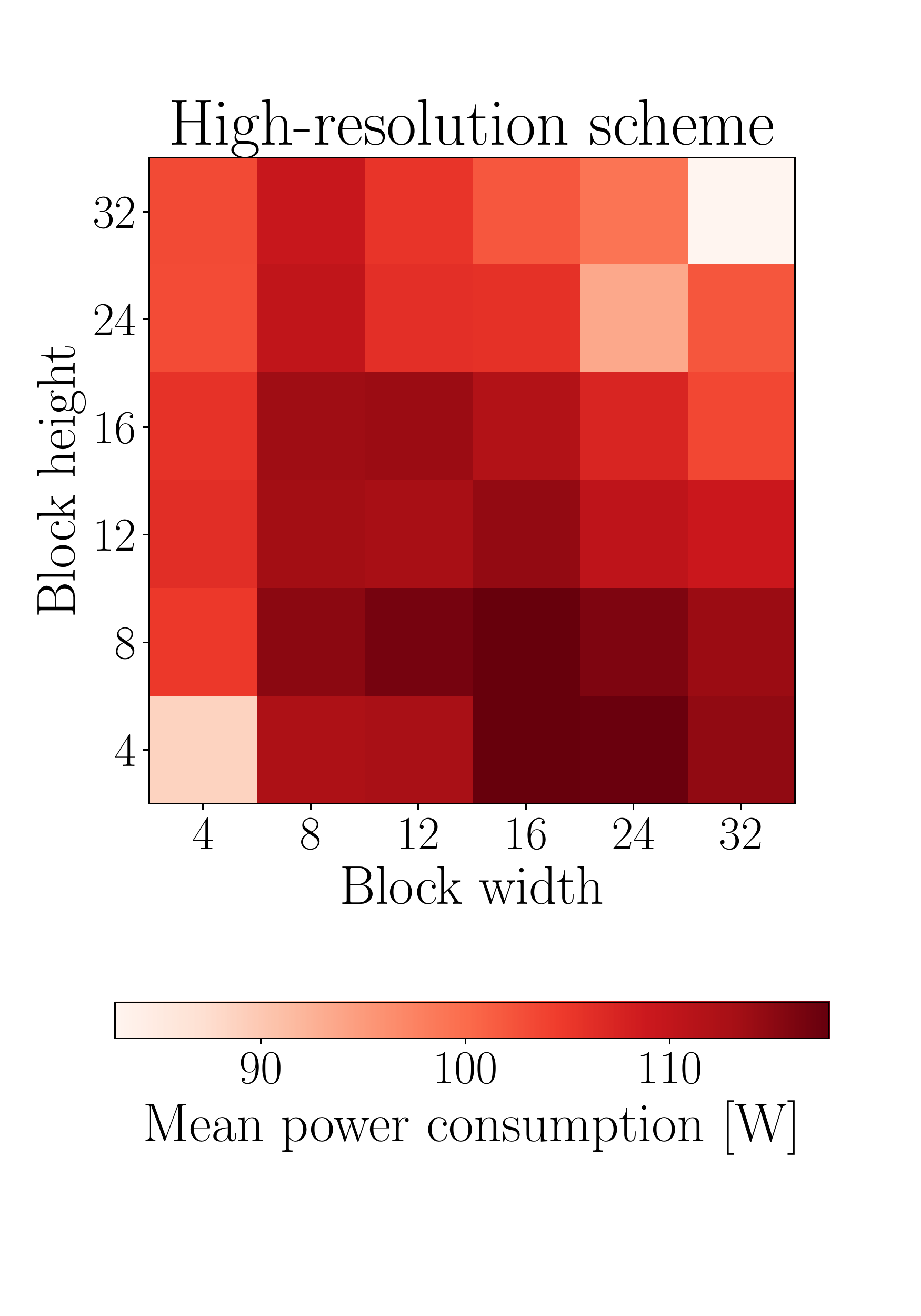}
    \hfill\null
    \\ \vspace{0.2cm}
    \hfill
    	\includegraphics[width=\figwbs\linewidth, trim=0.6cm 3.1cm 1.2cm 3.0cm, clip]{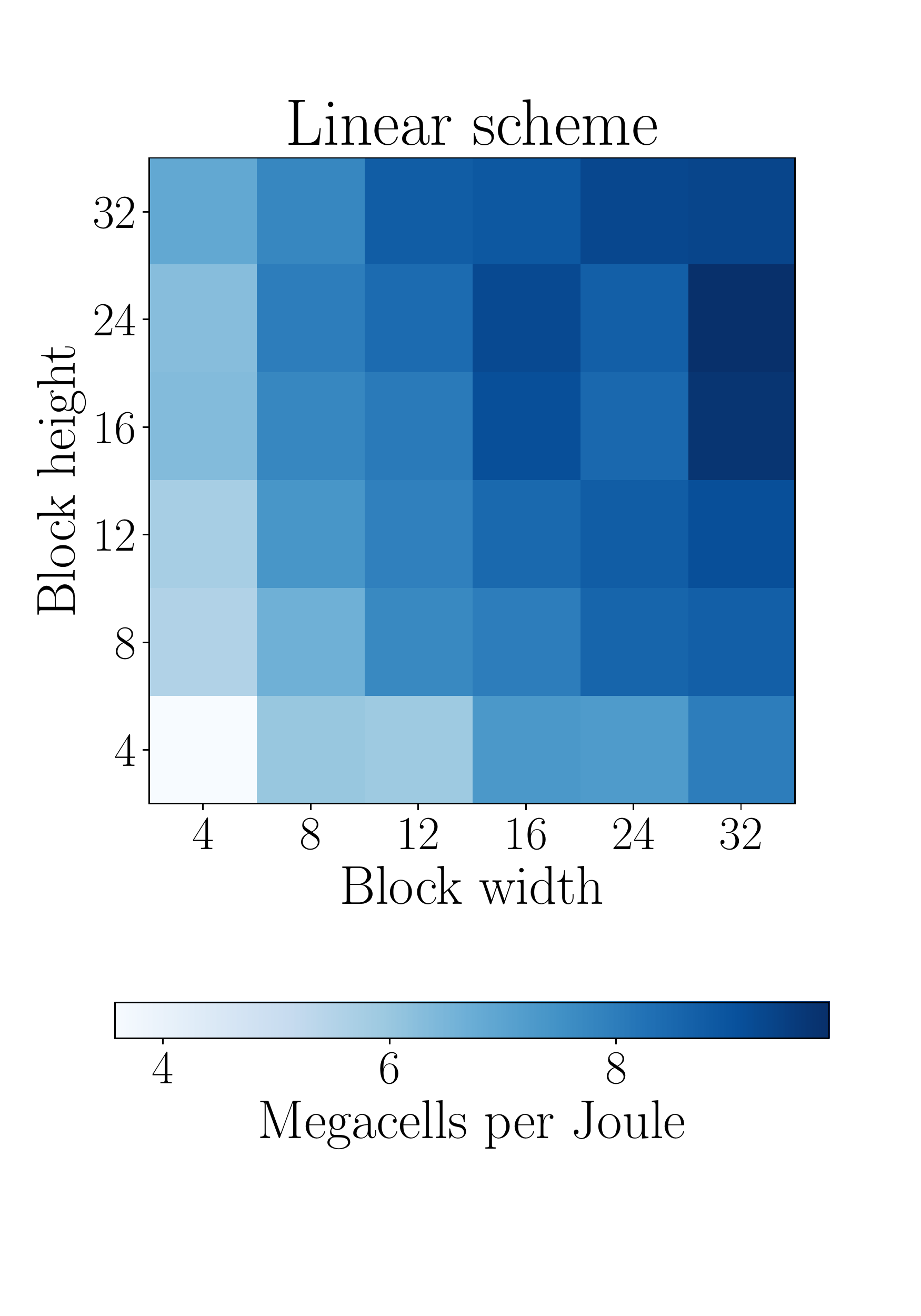}
    	\,
    	\includegraphics[width=\figwbs\linewidth, trim=0.6cm 3.1cm 1.2cm 3.0cm, clip]{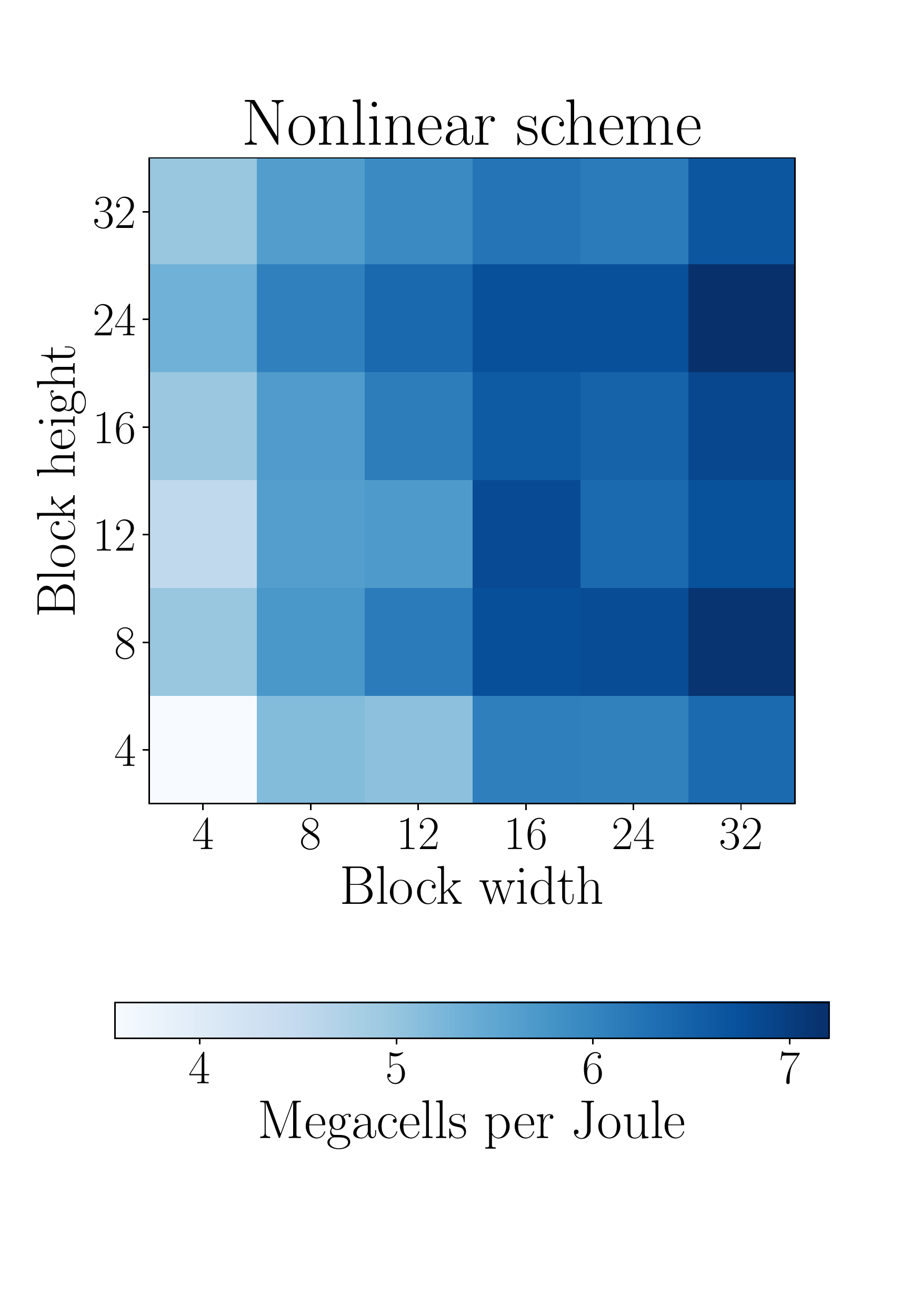}
        \,
    	\includegraphics[width=\figwbs\linewidth, trim=0.6cm 3.1cm 1.2cm 3.0cm, clip]{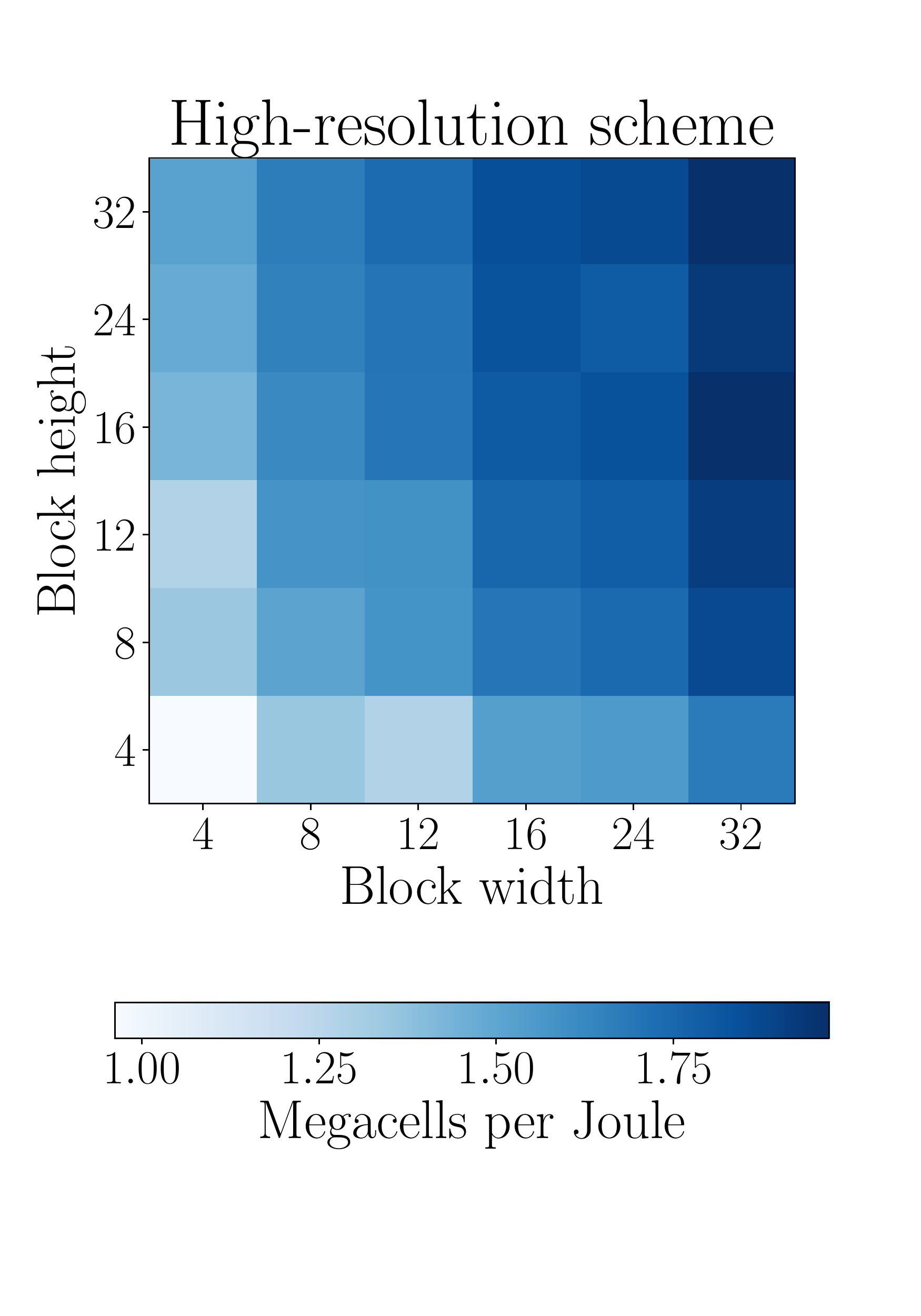}
    \hfill\null
    \caption{Computational performance (top), mean power consumption (middle), and energy efficiency (bottom) for the three numerical schemes for different block size configurations on the Tesla P100 GPU.
    Notice that all three measures change rapidly and unpredictably between the different configurations, and that there does not seem to be a direct translation between high computational performance and the mean power consumption.
    The bottom row shows that it is just as important to tune for optimal block size configuration when optimizing for energy efficiency as when optimizing for computational performance, and that these two objectives cannot be expected to be fulfilled simultaneously.}
    \label{fig:power_block_size}
    \end{center}
\end{figure}

Figure~\ref{fig:power_block_size} shows benchmark results for the three different schemes for a wide range of block size configurations.
The top row shows computational efficiency, in terms of computed megacells per second, and confirms how sensitive the performance for all three schemes are to different configurations, similarly to the results from Figure~\ref{fig:block_size}.
The optimal configurations with respect to computational efficiency are $(16, 16)$, $(16, 4)$ and $(32, 8)$ for the linear, nonlinear and high-resolution scheme, respectively.
The second row shows the mean power consumption for each execution.
By comparing the heat map of mean power usage with the computational efficiency, we see that high computational performance does not necessarily translate into high mean power consumption.
As a specific example, the most efficient configuration for the linear scheme has a lower mean energy consumption than half of the configurations.
The final row of Figure~\ref{fig:power_block_size} shows energy efficiency with respect to block size configuration, and again we see that the results do not follow an easily predictable pattern across the different block sizes.
Notice how the most energy efficient block sizes ($(32,24)$ for the linear and nonlinear scheme, and $(32,32)$ for the high-resolution scheme) can be recognized as configurations that have particularly low mean power consumption and are less than medium fast.
None of the most computational efficient configurations are among the very best configurations in terms of energy efficiency.
These results show that block tuning for computational efficiency and for energy efficiency are not the same.
However, because a performance tuned code has a lower time-to-solution, we can in general claim that performance tuning increases the power efficiency proportionally to performance.

\section{Summary}
\label{sec:summary}

We have presented our experiences from working with CUDA and OpenCL from Python for an extensive period of time\footnote{The authors have worked with GPU computing from C++ for over ten years, and over four years from Python}. 
Our conclusion is that using Python is as computationally efficient as C++ for our use, and we believe this to be true for many other application areas as well.
We have also benchmarked three different OpenCL codes, our ported code in CUDA, our optimized CUDA code, and finally our OpenCL code with optimizations found using the CUDA tools. 
Our results are shown for seven different GPUs, thus representing many of the GPU architectures in use today.
Finally, we have looked at the power consumption for all versions of the code and the potential for tuning for energy efficiency.

One of our most important findings is that working with Python is significantly faster than using C++, as we have previously done. We notice that the amount of research we are able to do in the same amount of time increases dramatically, and that the code quality is higher with fewer bugs and crashes. The combination of PyCUDA/PyOpenCL and the Jupyter Notebook led to a very productive development environment after addressing the initial crashes that forced us to reboot often. For our application area, the overhead of using Python becomes negligible with respect to performance.

The original motivation for using OpenCL was to support GPUs and similar architectures from multiple vendors. Our motivation for changing from OpenCL to CUDA was because of the better software ecosystem for CUDA, and we have been very happy with our decision. CUDA appears to be a much more stable and mature development ecosystem with better tools for development, debugging and profiling for our hardware.

We found it interesting that our initial port from OpenCL to CUDA imposed a performance penalty, due to different default compiler optimizations. 
Even though some authors have reported OpenCL to be slower than CUDA, we find no conclusive results that support this in general. The performance gain varied much more with the GPU being used than whether we used CUDA or OpenCL.
In most of our experiments, tuned versions of CUDA and OpenCL was found to have the same computational performance, with a few examples showing CUDA being faster than OpenCL.
Additionally, we found that the performance gain of a single optimization strategy will have vastly different effects on the run time for different GPUs. Even though we profiled and optimized mainly using a laptop GPU, the highest relative performance gain was for a server class and a desktop class GPU (the high-resolution scheme on the GTX780 and K20).

There does not seem to be any clear relationships between the power consumption when comparing different schemes, optimization levels, GPUs, and programming models.
When we consider power efficiency, we see that CUDA performs better than OpenCL for all tuned schemes on the Tesla V100 and GeForce 840M GPUs, whereas there are much smaller differences on the GeForce GTX780 and Tesla P100 GPUs.
When we examine the impact of performance tuning on power efficiency, there appears to be a systematic and clear relationship: A fast code is a power efficient code.

In terms of mean power consumption we find no clear relationships when comparing different schemes, optimization levels, GPUs, or programming models, but it seems to be just as important to consider block size configurations, as we measure a factor two difference between the lowest and highest mean power for different configurations.
We also see that optimizing block size for power efficiency is not necessarily the same as optimizing for computational efficiency.


\subsection*{Acknowledgement}
This work is supported by the Research Council of Norway through grant number 250935 (GPU Ocean).
The Tesla K20 computations were performed on resources provided by UNINETT Sigma2 -- the National Infrastructure for High Performance Computing and Data Storage in Norway under project number nn9550k.

\subsection*{Author Contributions}
The authors have all contributed to all aspects related to the work, including methodology, software, benchmarking and preparing the manuscript. The authors have contributed close to equally to most parts of the paper, but H.H.H. has had a lead role in the aspect directly related to power efficiency, and has had the role of lead and corresponding author.

\subsection*{Conflicts of Interest}
The authors declare no conflict of interest.
The funders had no role in the design of the study; in the collection, analyses, or interpretation of data; in the writing of the manuscript, or in the decision to publish the results.

\bibliography{references}

\end{document}